\documentclass[10pt,journal]{IEEEtranTCOM}

\usepackage[english]{babel}
\usepackage[usenames]{color}
\usepackage[cp1250]{inputenc}
\usepackage{amsfonts}
\usepackage{amsthm}
\usepackage{graphicx}
\usepackage{epsfig}
\usepackage{mathrsfs}
\usepackage{amsmath}
\usepackage{algorithm}
\usepackage{algorithmic}
\usepackage{hyperref}
\usepackage[table]{xcolor}

\theoremstyle{plain}

\begin{document}
\newcommand{\bea}{\begin{eqnarray}}
\newcommand{\eea}{\end{eqnarray}}
\newcommand{\be}{\begin{equation}}
\newcommand{\ee}{\end{equation}}
\newcommand{\beas}{\begin{eqnarray*}}
\newcommand{\eeas}{\end{eqnarray*}}
\newcommand{\bs}{\backslash}
\newcommand{\bc}{\begin{center}}
\newcommand{\ec}{\end{center}}
\def\SC {\mathscr{C}}

\title{Framework for liquid crystal based particle models}
\author{\IEEEauthorblockN{Jarek Duda}\\
\IEEEauthorblockA{Jagiellonian University, Cracow, Poland,
\emph{jaroslaw.duda@uj.edu.pl}}}
\maketitle

\begin{abstract}
Long-range e.g. Coulomb-like interactions for (quantized) topological charges are observed experimentally in liquid crystals, bringing open question this article is exploring: how far can we take this resemblance with particle physics? Uniaxial nematic liquid crystal of ellipsoid-like molecules can be represented using director field $\vec{n}(x)$ of unitary vectors. It has topological charge quantization: integrating field curvature over a closed surface $\mathcal{S}$, we get 3D winding number of $\mathcal{S}\to S^2$, which has to be integer - getting Gauss law with finally built-in missing charge quantization if interpreting field curvature as electric field. This article proposes a general mathematical framework \textit{LdGS}: combining Landau-de Gennes field with Skyrme kinetic term, to extend this similarity with particle physics to biaxial nematic, getting surprising agreement with the Standard Model. Specifically, recognising intrinsic twist of uniaxial nematic allows hedgehog configurations with one of 3 distinguishable axes: having the same topological charge, but different energy/mass - getting similarity with 3 leptons. Topological vortices correspond to quark strings building baryons and nuclei. Vacuum dynamics extends electromagnetism from 3D rotation dynamics, with Klein-Gordon-like equation for twists corresponding to quantum phase. Like in Einstein's teleparallelism we can add 4th time axis, extending vacuum dynamics to SO(1,3) Lorentz group by boosts, getting additional second set of Maxwell equations for GEM (gravitoelectromagnetism) approximation of general relativity.
\end{abstract}
\section{Introduction}
\begin{figure}[t!]
    \centering
        \includegraphics[width=9cm]{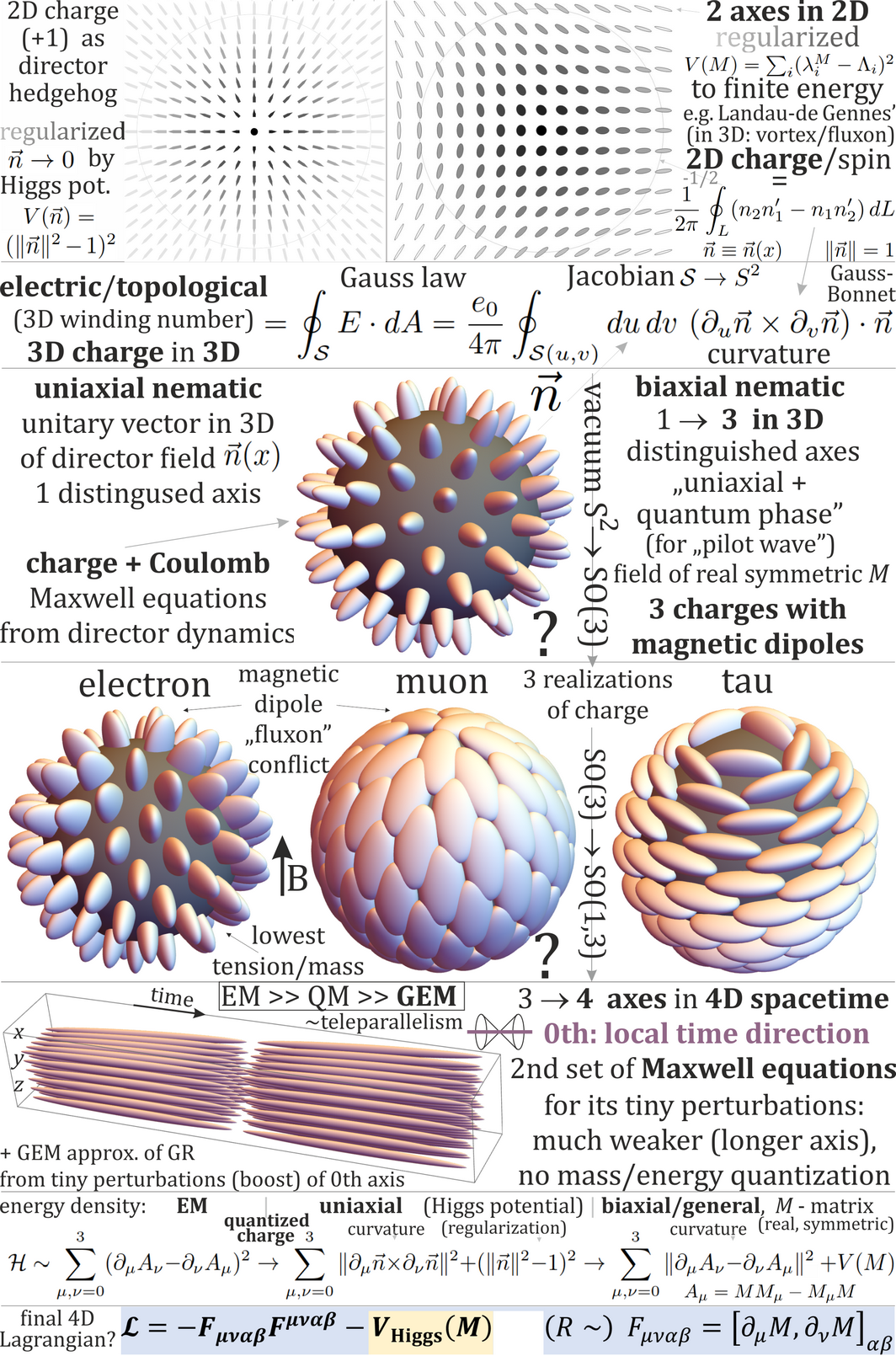}
        \caption{Unitary vector $\vec{n}\equiv \vec{n}(x)$ "director" field has \textbf{quantization of topological charge}, for which in liquid crystal experiments there were obtained \textbf{long-range interactions} - interpreting curvature as electric field, we can get \textbf{Maxwell equations} for their dynamics. Living in 3D suggests to extend it to 3 distinguishable axes representing rotating object - realized e.g. in biaxial nematic, getting \textbf{3 types of hedgehog} with the same charge, but different energy (like 3 leptons), also magnetic singularity due to the \href{https://en.wikipedia.org/wiki/Hairy_ball_theorem}{hairy ball theorem}. We model such ellipsoid field  like stress-energy tensor: with field of real symmetric matrices $M(x)\equiv M=ODO^T$ (orthogonal $OO^T=I$, $D=\textrm{diag}(\lambda_1,\lambda_2,\lambda_3)$), which \textbf{prefers some shape as set of eigenvalues due to Higgs-like potential} e.g. $V(M)=\sum_i (\lambda_i-\Lambda_i)^2$ for some fixed model parameters: $(\Lambda_i)$ - allowing to regularize singularity (discontinuity in the center) to finite energy as in top diagrams.  Reminding that we live in 4D spacetime, like in Einstein's teleparallelism, suggests to add 0th time axis as the longest: with \textbf{$\Lambda_0 (\textrm{gravity}) >> \Lambda_1 (\textrm{EM}) >>\Lambda_1 (\textrm{QM}) > \Lambda_0 (0)$}, having the strongest tendency to be aligned in parallel, acting as local time direction: central axis of light cone.  Mass/energy should enforce tiny perturbations as boost curvature of this time axis, with dynamics given by \textbf{second set of Maxwell equations for GEM} approximation of the general relativity, for boosts in vacuum extension from SO(3) to SO(1,3) Lorentz group.}
        \label{intr}
\end{figure}

While particles are usually treated in perturbative way, this is only effective approximation of nonperturbative QFT: asking for field configurations, to finally consider their Feynman ensembles in 2nd quantization. This fundamental question is now asked nearly only for baryons in lattice QCD~\cite{lattice}, while we should try to understand field structures of all particles.

Classical electromagnetic field naively has two crucial issues we should repair before considering its Feynman ensemble:
\begin{enumerate}
  \item missing \textbf{charge quantization}: Gauss law should only return integer multiplicities of $e$ (+confined fractional),
  \item missing \textbf{regularization}: infinite energy of electric field of charge, bounded by 511keV released in annihilation.
\end{enumerate}

In liquid crystals there are experimentally realized quantized topological charges with long-range interactions, e.g. resembling quadrupole-quadrupole~\cite{lq1}, dipole-dipole~\cite{lq2}, Coulomb~(\cite{lq3,lq33}) or even stronger~\cite{lq4} interactions - suggesting resolution to both problems on classical field level, for example using Faber's approach~(\cite{faber1,faber15,faber2,faber3,faber4}) this article extends on - summarized in Figure \ref{intr} (gathered  materials\footnote{\url{https://github.com/JarekDuda/liquid-crystals-particle-models/}}):

\begin{figure*}[t!]
    \centering
        \includegraphics[width=18cm]{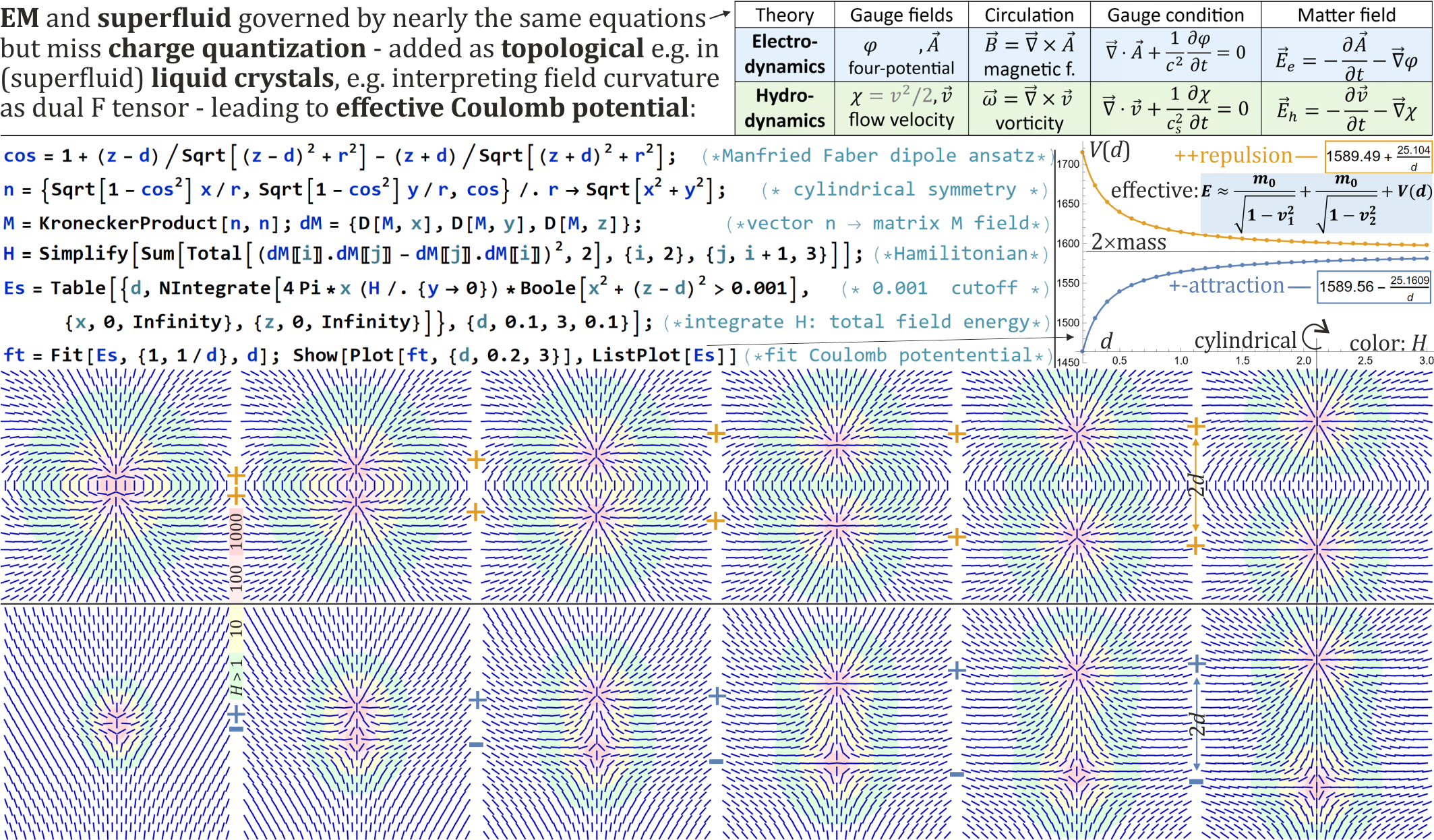}
        \caption{EM-hydrodynamics analogy~\cite{EMh} (top) missing charge quantization in Gauss law, added as topological - below calculation of Coulomb effective potential with shown  Mathematica source (extended in GitHub). For "+-" or "++" pair of topological charges, there was postulated ansatz~\cite{faber15}: cylindrically symmetric configuration of unitary vector field $\vec{n}\equiv \vec{n}(x)$ in agreement with electric field of two elementary charges (in GitHub verified satisfaction of variation equation). Such various distance configurations are shown with visualized $H$ density. Seen as uniaxial nematic it corresponds to $M=nn^T$ matrix field, as discussed here with static energy density $H=\sum_{1\leq i<j\leq 3} \|[\partial_i M,\partial_j M]\|^2$. Integrating this energy density with cutoff around two singularities, there was numerically obtained $E(d)\approx 1590 \pm 25/d$ distance-energy dependence as in Coulomb law (shown values and fit). Finally the two singularities are to be regularized with Higgs-like potential e.g. $(1-\|n\|^2)^2$, due to Lorentz invariance leading to $m_0 \to m_0/\sqrt{1-v^2}$ energy scaling. To avoid infinite energy of singularities in charges there was used cutoff above, which finally should be replaced with regularization by Higgs-like potential - as discussed and calculated in \cite{faber4}, leading to deformations of Coulomb interaction in tiny distances in agreement with the running coupling effect. }
        \label{coulomb}
\end{figure*}

\begin{enumerate}
  \item Interpret curvature of e.g. field vector $\vec{n}$ as electric field, making \textbf{Gauss law counts its topological charge},
  \item \textbf{Use Higgs-like potential} e.g. $V=(\|\vec{n}\|^2-1)^2$, allowing for e.g. $\vec{n}\to 0$ regularization of  singularities.
\end{enumerate}

Then we can consider Feynman ensembles of such fields, allowing to resolve infinite energy issue before 2nd quantization: through field regularization corresponding to later renormalization, understand the differences between e.g. electron's field configuration and of (infinite energy) perfect point charge.

Liquid crystals use ellipsoid-like molecules, which if cylindrically symmetric (uniaxial nematic) can be represented with director field $\vec{n}\equiv \vec{n}(x)$ of unitary vectors, this way allowing for (quanitzed) topological charges e.g. hedgehog-like configurations. They get long-range e.g. \textbf{Coulomb-like interaction} as in Figure \ref{coulomb}: total energy of the field (as integrated energy density: Hamiltonian) for two charges in various distances behaves as in Coulomb potential.

Such 3D topological charge as 3D winding number~\cite{wind} of $\vec{n}$ restricted to $\mathcal{S}\to S^2$, can be calculated by integrating over closed surface $\mathcal{S}$ the Jacobian of this function - which turns out curvature of this field. Therefore, interpreting (e.g. vector) field curvature as electric field, we get \textbf{Gauss law with built-in charge quantization as topological} (\href{https://en.wikipedia.org/wiki/Gauss\%E2\%80\%93Bonnet_theorem}{Gauss-Bonnet theorem}).

\begin{figure}[t!]
    \centering
        \includegraphics[width=85mm]{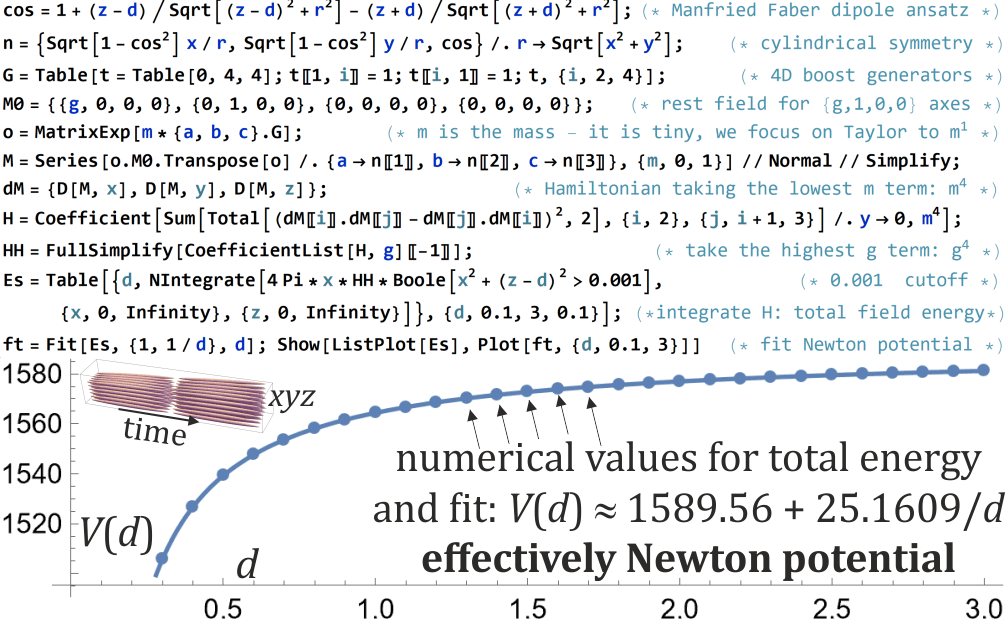}
        \caption{Example of simplified calculation of Newton effective potential (in GitHub) - analogously as in Fig. \ref{coulomb}, but this time instead of large spatial rotations, using tiny boosts of 0th time axis for gravity (no mass quantization). Spherically symmetric curvature sources would have increased energy with reduced distance, hence to get attraction there was used dipole ansatz for microscopic scenarios like pair creation, hopefully averaging to spherically symmetric gravity. Final calculations will need further work.}
        \label{newton}
\end{figure}

\begin{figure}[t!]
    \centering
        \includegraphics[width=9.2cm]{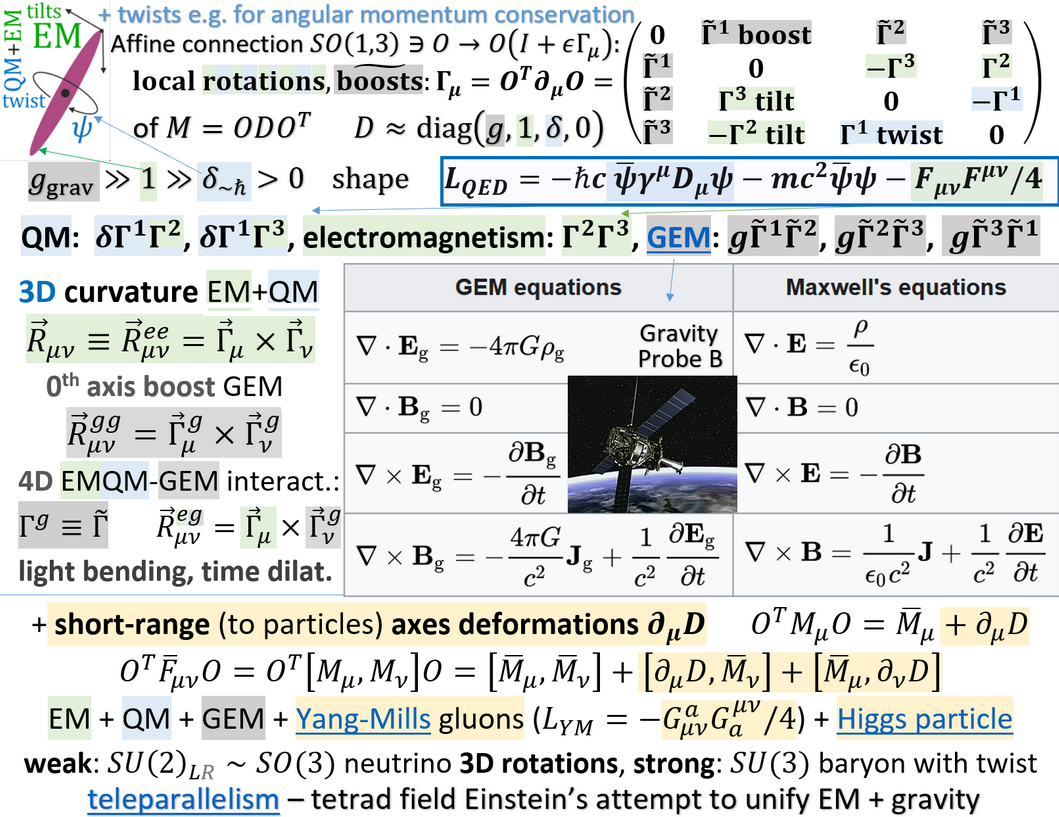}
        \caption{Summary of interactions by SO(1,3) dynamics of 4 orthogonal axes like in \href{https://en.wikipedia.org/wiki/Teleparallelism}{Einstein's teleparallelism}, but of different lengths of axes: $g \gg 1 \gg \delta >0$. The 0th axis of length $g\sim 10^{10}$ is local time direction requiring high energy to change, its boost dynamics recreates \href{https://en.wikipedia.org/wiki/Gravitoelectromagnetism}{gravitomagnetism} (GEM) Maxwell equations of gravity confirmed by \href{https://en.wikipedia.org/wiki/Gravity_Probe_B}{Gravity Probe B} (diagrams) - minimal Lorentz invariant extension of Newton force in analogy to Coulomb, used as approximation of General Relativity. $S^2$ dynamics of 1st axis having length 1 gives EM Maxwell equations, and electric charge quantization as topological. Nonzero 2nd axis of length $\delta\sim 10^{-10}$ allows for tiny energy contributions of U(1) twists of 1st axis, as quantum phase in QED Lagrangian. EM-GEM interaction e.g. slows down EM propagation in gravitational field - leading to gravitational time dilation, and light lensing through Fermat principle. Additionally, there are degrees of freedom deforming these $(g,1,\delta,0)$ eigenvalues preferred by Higgs-like potential, activated mainly near particles for regularization, which resemble e.g. \href{https://en.wikipedia.org/wiki/Yang\%E2\%80\%93Mills_theory}{Yang-Mills} Lagrangian contributions.}
        \label{GEM}
\end{figure}

The center of such quantized topological charge naively has field discontinuity, which would mean infinite energy - like of electric field around point charge. To prevent it, we could make a cutoff as in Fig. \ref{coulomb}, in liquid crystals we can imagine there is no molecule in the center of e.g. hedgehog configuration. However, for a field there should be value everywhere - we need to \textbf{regularize} it, deform  to finite energy, e.g. at most 511 keVs for electron - released in annihilation. There can be used $V(\vec{n})=(\|\vec{n}\|^2-1)^2$ Higgs potential: preferring unitary vectors, also allowing to deform e.g. to $\vec{n}= 0$ in the center of singularity to prevent infinity. Massless dynamics of this vacuum (Goldstone bosons) can be  chosen to resemble electromagnetism by interpreting curvature as electric field. Experimental consequence of such regularization to finite energy is deformation of Coulomb interaction in tiny distances, which agrees with known \textbf{running coupling} effect~\cite{faber4}. For regularization of a more general field, we need potential with topologically nontrivial minimum, e.g. $S^2$: $\|\vec{n}\|=1$ for uniaxal nematic, SO(3) for biaxial nematic, like in Landau-de Gennes model~\cite{gennes}, further extended to SO(1,3) to add gravity as in Fig. \ref{GEM}, getting field of 4 axes as in Einstein's teleparallelism~\cite{tele}.

Above director field $\vec{n}$ does not recognize twist of this vector, hence not need to conserve such angular momentum. To repair it, we can use generic objects in 3D with 3 distinguishable axes, SO(3) rotations - like molecules in experimentally challenging biaxial nematic liquid crystals~(\cite{gennes,biax,bi}). Similarly to Landau-de Gennes model, we will represent such unknown rotation using field of real symmetric matrices $M(x)\equiv M=ODO^T$ (orthogonal $OO^T=I$, $D=\textrm{diag}(\lambda_1,\lambda_2,\lambda_3)$): which due to potential prefers some fixed set of eigenvalues $\Lambda_1>\Lambda_2>\Lambda_3$, but allows for their $(\lambda_i)$ deformation/regularization to prevent infinite energy in singularity. E.g. using $V(M)=\sum_i (\lambda_i-\Lambda_i)^2$ Higgs-like potential, as in top-right diagram in Fig. \ref{intr}. Instead of integer, this 2D configuration has 2D topological charge +1/2 due to symmetry: that ellipsoid rotated by $\pi$ is the same ellipsoid, what also agrees with quantum rotation operator: "rotating spin $s$ particle by $\phi$ angle, rotates phase by $s \phi$" - suggesting to \textbf{interpret 2D topological charge as spin}, \textbf{3D as electric charge} and use symmetric $\vec{n}\equiv -\vec{n}$ field to allow spin 1/2. Additionally, fluxons in superconductor are well known 2D topological charges, like spin associated with magnetic field. Also, voricity works analogously to magnetic field e.g. in Aharonov-Bohm~(\cite{berry,berry1}) and hydrodynamical Zeeman effect~\cite{c4}. Symmetry $\vec{n}\equiv -\vec{n}$ also prevents domain walls, not observed in particle physics.

\begin{figure}[t!]
    \centering
        \includegraphics[width=9 cm]{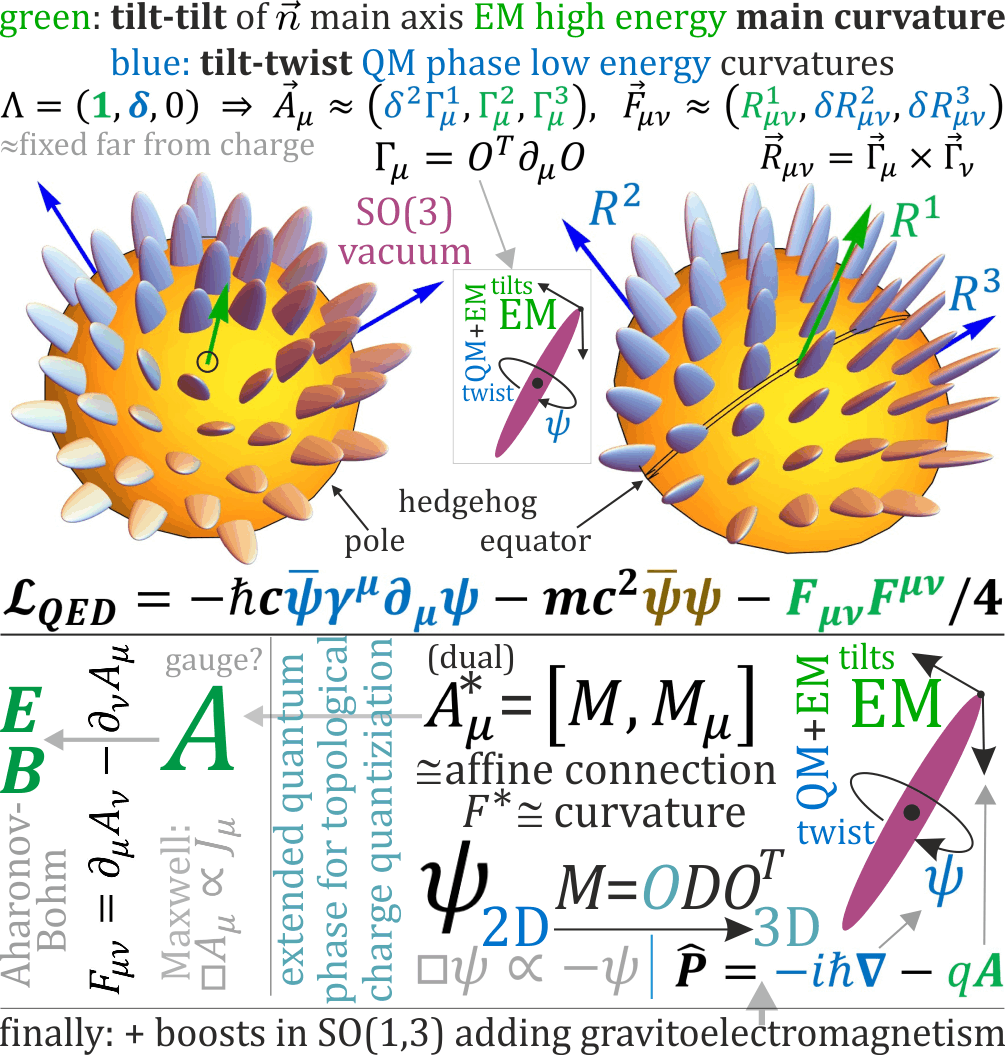}
        \caption{\textbf{Top}: parts of hedgehog of the longest $\vec{n}$ \textbf{main axis} of biaxial-nematic-like field. This axis can \textbf{tilt} in two directions already in uniaxial nematic, in biaxial there are additionally recognized its \textbf{twists} - here tilts correspond to high energy EM dynamics, twists to low energy QM phase dynamics. Such local rotation is affine connection $\Gamma_\mu=O^T (\partial_\mu O)$: antisymmetric matrix we can interpret as local rotation vector $\vec{\Gamma}_\mu := \left((\Gamma_{\mu})_{32},(\Gamma_{\mu})_{13},(\Gamma_{\mu})_{21}\right)$. It corresponds to $A_\mu$ four-vector weighted with shape $\lambda_i\approx \Lambda_i$ (far from charge fixed by potential e.g. $V=\sum_i (\lambda_i -\Lambda_i)^2$) distinguishing high energy tilts from low energy twists. Curvatures $\vec{R}_{\mu\nu}=\vec{\Gamma}_\mu \times \vec{\Gamma}_\nu$ after weighting with shape become dual $F^*_{\mu\nu}$ tensor: containing high energy tilt-tilt component $R^1_{\mu\nu}$ corresponding to EM, and low energy tilt-twist $R^2_{\mu\nu},R^3_{\mu\nu}$ corresponding to QM phase like in QED-like Lagrangian. \textbf{Bottom}: due to Aharonov-Bohm-like arguments, there is belief that $A$ four-vector is more fundamental than $E,B$ fields, however, it leaves gauge freedom. It also allows for non-integer charges like half-electron - to prevent that, \textbf{there is postulated more fundamental field $M$ which (quantized) topological charge is calculated in Gauss law}. This deeper field can be seen as extended quantum phase: from low energetic evolution of U(1) quantum phase (twists), to SO(3) evolution (+tilts) including also electromagnetism with built-in (topological) charge quantization, getting natural EM+QM unification (+GEM with 4D field, SO(1,3) vacuum).        }
        \label{diagram}
\end{figure}

Distinguishing two types of rotation (of $\Lambda$ shape): twist of biaxial nematic, and two tilts as in Fig. \ref{diagram}, will allow to assign them different energy scales. Going to 4D, we can imagine additional much longer 0th time axis undergoing tiny perturbations - naively rotations of SO(4), but Lorentz invariance suggests to use SO(1,3) with boosts instead. Finally the discussed approach allows to \textbf{unify 3 types of vacuum dynamics} (far from particles/singularises):
\begin{itemize}
  \item \textbf{electromagnetism} (EM) of relatively high energy - governed by Maxwell (wave-like) equations, corresponding to tilts, already in uniaxial nematics (e.g. Coulomb: \cite{lq3}),
  \item \textbf{quantum phase} ($\arg(\psi)$) evolution of much lower energies ($\hbar c$ in QED Lagrangian) - corresponding to twists, governed by Klein-Gordon-like (wave-like) equation,
  \item \textbf{gravitoelectromagnetism} (GEM)\footnote{GEM: https://en.wikipedia.org/wiki/Gravitoelectromagnetism} approximation of general relativity - 2nd set of (wave-like) Maxwell equations for tiny perturbations (boosts) of 0th time axis.
\end{itemize}

Such field of 3 distinguishable axes allows to construct hedgehog-like configuration of one of 3 axes as in Fig. \ref{intr} - they have the same 3D topological charge acting as electric charge, but require different regularization/deformation - should have different mass/energy, resembling 3 leptons. Additionally, the hairy ball theorem~\cite{hairy} says that we cannot continuously align such axes on the sphere - requiring additional spin-like singularities resembling fluxons, which should correspond to magnetic dipole moment of leptons. In particle physics three families are very common: for leptons, neutrinos, quarks - here as just consequence of living in 3D.

\begin{figure}[t!]
    \centering
        \includegraphics[width=9cm]{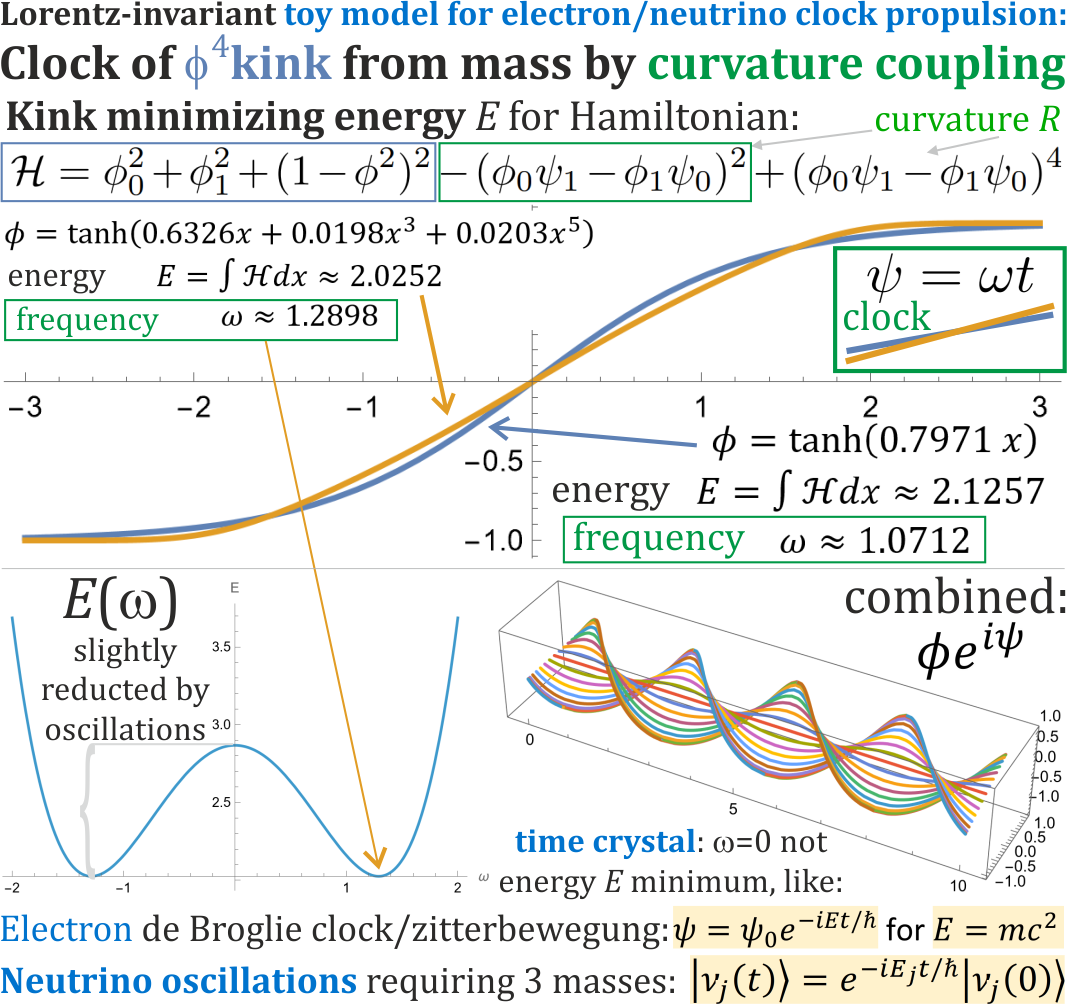}
        \caption{While usually energy minimization removes motion as kinetic energy, somehow resting neutrino oscillates, also electron as de Broglie clock and having angular momentum (of field not point), making them \href{https://en.wikipedia.org/wiki/Time_crystal}{time crystals} - originally as oscillating in the lowest energy, also observed in liquid crystals~\cite{liqtc}. In relativistic quantum mechanics we need $E=mc^2$ in $\psi\sim \exp(-iEt/\hbar)$, so their oscillations are literally propelled by mass. In nonperturbative field theories mass is integral of Hamiltonian - we need its negative terms for clock propulsion, which in discussed approach come from $F_{\alpha\beta\mu\nu}F^{\alpha\beta\mu\nu}$ Lagrangian term with 4 indexes, as $F$ is dual curvature requiring 4 indexes, which transformed to Hamiltonian from spacetime signature gets also negative terms as in further Fig. \ref{Fmunu}. The diagram shows simplified 1+1D toy-model from \cite{timecryst}, in which including negative squared curvature Lorentz-invariant term in Hamiltonian of popular $\phi^4$ model, particle as kink can slightly reduce energy by activating oscillations. }
        \label{kink}
\end{figure}

The difference between uniaxial and biaxial nematic can be imagined as recognizing intrinsic rotation (referred as twist) of elongated molecule - here adding to electromagnetism (referred as tilts) single low energy vacuum degree of freedom (crucial for angular momentum conservation), which seems to correspond to quantum phase, pilot wave. Quantum mechanical phase evolution $\exp(-iEt/\hbar)$ in relativistic e.g. Dirac equation requires $E=mc^2$, leading to caused by mass itself $\omega=mc^2/\hbar$ frequency oscillations already for resting particles, for example of neutrinos between flavors. For electron it is $\sim 10^{21}$ Hz and was originally postulated by Louis de Broglie, is also called \href{https://en.wikipedia.org/wiki/Zitterbewegung}{Zitterbewegung}, and has multiple experimental confirmations, e.g. direct~\cite{clock}, or in Bose-Einstein condensate analogs~\cite{zitt}. Hence the discussed corresponding configurations should enforce such periodic process, like spin precession~\cite{spin} or rather rotation (twist) of this additional degree of freedom - leading to "pilot wave" coupled with such electron. For 3D case there is obtained Klein-Gordon-like equation, but missing gravitational mass - added in 4D considerations: due to spacetime signature, there appear subtle negative energy Hamiltonian terms (shown further in Fig. \ref{Fmunu}), exactly as required to propel such oscillations. Its 1+1D toy model from \cite{timecryst} is shown in Fig. \ref{kink}.

Analogous view on wave-particle duality has also allowed for experimental realizations of hydrodynamical analogs of many quantum phenomena with \textbf{hydrodynamical wave-particle duality objects}: walking droplets. For example double slit interference~\cite{c1} (corpuscle travels one trajectory, its coupled wave travels all - affecting corpuscle trajectory), unpredictable tunneling~\cite{c2} (depending on complex history of the field), Landau orbit quantization~\cite{c3} (coupled wave has to become standing wave for resonance condition in analogy to stationary Schr\"{o}dinger equation for orbital quantization), Zeeman-like splitting~\cite{c4} of such quantized orbits (using Coriolis force as analogue of Lorentz force, vorticity as magnetic field~(\cite{berry})), double quantization~\cite{c5} (in analogy to $(n,l)$ for atomic orbitals), recreating quantum statistics with averaged trajectories~\cite{c6}, Elitzur-Vaidman bomb testing~\cite{EV}, or Bell violation~\cite{Bellh}. There are also known hydrodynamical analogs for Casimir~\cite{casimir} and Aharonov-Bohm effects~(\cite{berry,berry1}). For fluxons as 2D topological charges in superconductor there was experimentally realized e.g. interference~\cite{f1}, tunneling~\cite{f2} and Aharonov-Bohm~\cite{f3} effect.

Maximal Entropy Random Walk (MERW) also suggests quantum-like statistics for objects undergoing complex dynamics. Standard diffusion models turn out to only approximate the (Jaynes) maximal entropy principle, necessary for statistical physics models - lacking Anderson-like localization, observed also for neutrons~\cite{neutron}. MERW finally does this optimization, getting stationary probability distribution exactly as quantum ground state, with its localization properties (\cite{merwprl, my, myphd,cond}). E.g. for $[0,1]$ range standard diffusion predicts uniform $\rho=1$ stationary probability distribution, while QM and MERW predict localized $\rho\propto \sin^2$.

As we live in 4D spacetime, it is natural to extend from 3 to 4 distinguishable axes by just going from $3\times 3$ to $4\times 4$ real symmetric matrix field $M$ - like the stress-energy tensor, for which $M$ might be microscopic extension. This way the field recognizes not only SO(3) rotations, but also boosts going to SO(1,3) vacuum, what is required for Lorentz invariance. The 0th axis should be the longest - having the strongest tendency to align in nearly parallel way. This way dynamics of its tiny perturbations (boosts) is governed by additional set of Maxwell equations - with goal to obtain e.g. GEM: confirmed by Gravity Probe B approximation of general relativity. Such tiny perturbation/spatial curvature can be caused e.g. by EM-GEM interaction or activating potential to give particles also gravitational mass. Slowing down of EM propagation through EM-GEM interaction could explain gravitational time dilation and light lensing~\cite{dicke}. In contrast to charge corresponding to complete spherical angle, this time we have only tiny curvatures - there is no mass quantization.

\begin{figure}[t!]
    \centering
        \includegraphics[width=9cm]{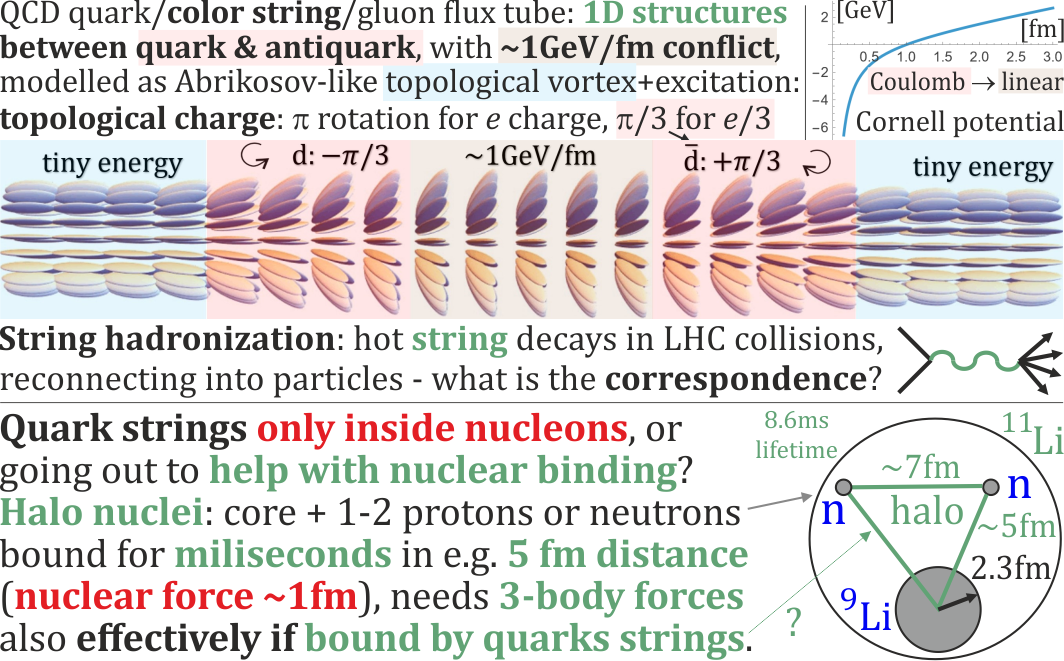}
        \caption{The central object of QCD is quark string, also called color or gluon flux tube. It connects quark-antiquark pairs, and has asymptotically linear $\sim 1$ GeV/fm energy density, confining the quarks - which in pairs can be created/annihilated on such string. This 1D structure is very stable, and nonperturbatively modeled as topological Abrikosov vortices~\cite{string}, fluxons, in liquid crystals called disclinations. To add \textbf{quarks as fractional charge excitations of quark string}, interpreting electric charge as topological here, we can do it by inward/outward field rotation: by $\pi$ would be elementary charge, hence for quarks there should be used fractional - as in above diagram, leading to conflict between them, asymptotically with linear energy/length as required for widely used \href{https://en.wikipedia.org/wiki/Cornell\_potential}{Cornell potential}. Gauss law for a region cutting such string, has (regularized) singularity in this point. As shown further in Fig. \ref{baryon}, \ref{part}, tendency to make such rotation is required by suggested baryon structure. Additionally, basic hadronization models used e.g. in LHC collider (\url{http://www.scholarpedia.org/article/Parton_shower_Monte_Carlo_event_generators}) is \textbf{string hadronization}~\cite{hadronization} - assuming formation of such hot string in collision, and analyzing results of its decay through reconnections. 
        Therefore, to understand field configurations of particles, we should \textbf{search for correspondence between topological vortex decay, and observed formed in collisions} - constraining toward the discussed approach. Basic objects made of 1D structures like quark strings are knots, also proposed for exotic particles~\cite{glue1,glue2,parknot}, but should be created in LHC collisions - we present arguments that these knots are just baryons/nuclei, e.g. \href{https://en.wikipedia.org/wiki/Halo_nucleus}{halo nuclei} with 1-2 nucleons bound for milliseconds in distance much larger than of nuclear force, requiring 3-body forces, like if being bound by 1D quark string.}
        \label{string}
\end{figure}

In liquid crystals, superfluids, superconductors there are also unavoidable 1D topological structures, called Abrikosov's vortex, fluxon, disclination. Searching for its correspondence in particle physics, the only candidate seem quark strings being at heart of QCD, briefly summarized in Fig. \ref{string}. They are believed to be decaying during string hadronization process in colliders like LHC, simplifying the task to search for correspondence between such results and decay of topological vortices. As discussed further, it automatically leads to looking perfect at least qualitative agreement.

Like electromagnetism, the discussed approach is viscosity-free, hence complete experimental realizations would require e.g. superfluid like in famous Volovik "Universe in a helium droplet" book~\cite{volovik}. However, simplified experimental settings could allow to get some interesting correspondence, like vortices going out of biaxial nematic topological charge due to the hairy ball theorem (no spin-less charge).

Related skyrmion models~(\cite{manton,skyrm}) use similar 4th order kinetic term, also aiming particle correspondence - mainly nuclei, instead of electric charge interpreting topological charge as baryon number. They lack long-range EM interaction due to potential with single minimum, repaired here with Higgs-like. Instead of electric charge conservation, they cannot violate the baryon number - what is questioned e.g. due to lack of Gauss law for baryon number, and violation required e.g. in baryogenesis (creation of more baryons than antibaryons) or Hawking radiation (massless from originally baryons).

\begin{figure}[t!]
    \centering
        \includegraphics[width=9cm]{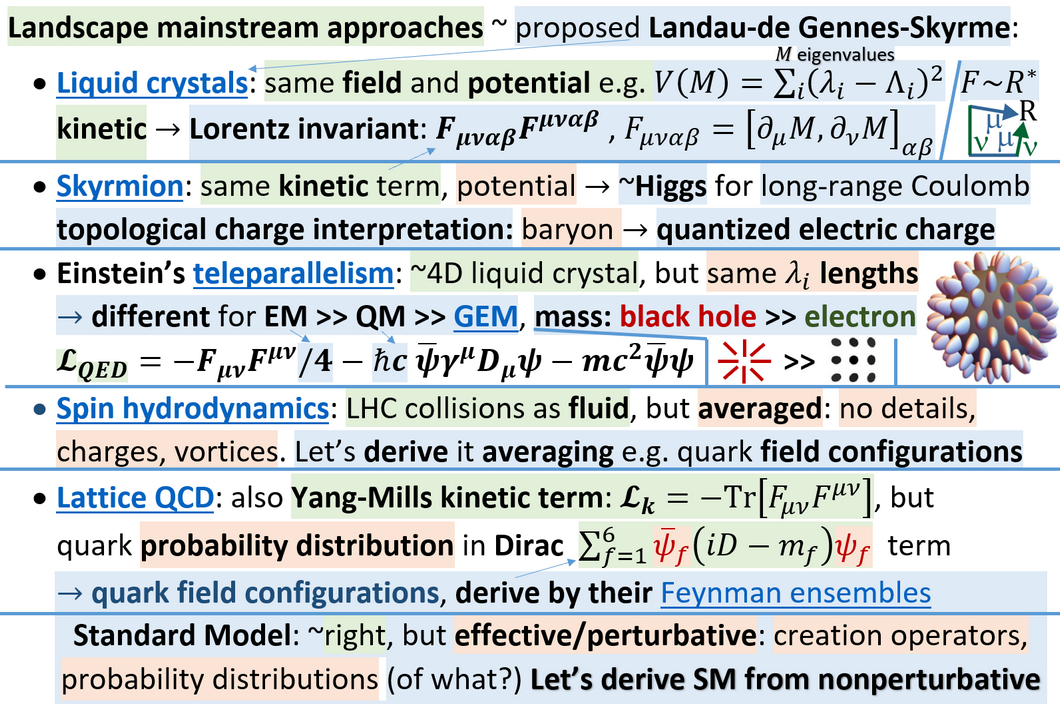}
        \caption{Summary of landscape of approaches related to discussed LdGS. \textbf{Liquid crystal} \href{https://en.wikipedia.org/wiki/Landau\%E2\%80\%93de_Gennes_theory}
        {\textbf{Landau-de Gennes model}} uses analogous field and potential, but lacks Lorentz invariance - repaired by replacing kinetic term with EM/\href{https://en.wikipedia.org/wiki/Yang\%E2\%80\%93Mills_theory}{Yang-Mills}/Skyrme-like. \href{https://en.wikipedia.org/wiki/Skyrmion}{\textbf{Skyrmion}} models interpret topological charge as baryon number - forbidding its violation (required e.g. by \href{https://en.wikipedia.org/wiki/Baryogenesis}{baryogenesis} and \href{https://en.wikipedia.org/wiki/Hawking\_radiation}{Hawking radiation}), and missing charges - we repair by switching to its interpretation as quantized electric charge, also changing potential to LdG/Higgs-like for long-range interactions. \href{https://en.wikipedia.org/wiki/Teleparallelism}{Einsten's \textbf{teleparallelism}} uses 4 orthonormal vector fields forming tetrad, like 4D liquid crystal here. However, its axes have same lengths, making energy of e.g. hedgehog in space (electron) and in time (black hole) similar, while they should be very different. \textbf{Spin hydrodynamics}~\cite{spinhydr} is also fluid-like for LHC collisions, but neglecting details like charges/vortices - we would like to derive by averaging over such details. \href{https://en.wikipedia.org/wiki/Lattice_QCD}{\textbf{Lattice QCD}} is one of the most fundamental approaches used mainstream, however, it still uses probability distributions of quarks, as in effective statistical physics models. Finally, the \textbf{Standard Model} provides good experimental agreement, but is perburbative/effective - beside probability distributions uses abstract creation operators, without asking what field configurations they describe. We would like to derive it from more fundamental model of lower number of parameters. }
        \label{landscape}
\end{figure}

While the presented general view was already discussed by the author~(the first version of \cite{my}, \cite{mytop}), this article finally introduces looking proper mathematical framework. The current version is updated work in progress, planned to be further extended in the future. Connections with related approaches are shown in Fig. \ref{landscape}. The main goal is to derive the Standard Model as effective, from Feynman ensembles of more fundamental nonperturbative LdGS model with concrete field configurations of  particles (instead of abstract creation operators) and without probability distributions. Such deeper level should allow to reduce the number of parameter from $\approx 30$ to a few, deriving relations between them like masses of particles by integrating Hamiltonian, also resolving issues like unification with gravity.

\section{General quantitative framework}\label{sec2}
This main Section first introduces to  electromagnetism with built-in charge quantization and regularization using director field in analogy to Faber's approach~(\cite{faber1,faber15,faber2,faber3,faber4}). Further there is discussed generalization to Landau-de Gennes biaxial nematic case using field of real symmetric matrices (like stress-energy tensor) preferring some shape as fixed $(\Lambda_i)$ set of eigenvalues, and then to 4D case as in Einstein's teleparallelism.
\subsection{Gauss law with built-in (topological) charge quantization}
Imagine a continuous (director) field of unitary vectors $\vec{n}:\mathbb{R}^d \to S^{d-1}$. Restricting it to a closed surface $\mathcal{S}\subset \mathbb{R}^d$ gives $\vec{n}:\mathcal{S}\to S^{d-1}$ function, which has some integer number of coverings/windings of this sphere - called topological charge.

This (generalized) winding number, multiplied by sphere area, can be obtained by integrating Jacobian (as determinant of Jacobian matrix) over this closed surface - we would like to interpret this Jacobian e.g. as electric field, making that Gauss law counts this winding number - getting missing built-in charge quantization as topological charge.

In 2D case, analogous e.g. to argument principle in complex analysis, integrating derivative of angle over loop gives $2\pi$ times topological charge ($\vec{n}' = d \vec{n}/dL$, $\vec{n}=(n_1,n_2)$):
$$\textrm{2D topological charge}=\frac{1}{2\pi}\oint_L (n_2 n'_1-n_1 n'_2)\, dL$$
This way we e.g. get quantization of magnetic field in superconductors as fluxon/Abrikosov vortex~\cite{fluxon} being 2D topological charge, which also resembles spin as in quantum rotation operator: saying that rotating spin $s$ particle by $\varphi$ angle rotates quantum phase by $s \varphi$. Observe that, as in top of Fig. \ref{intr}, with liquid crystals we can get spin 1/2 this way, as rotating by $\pi$ radians we get the same ellipsoid due to $\vec{n}\equiv -\vec{n}$ symmetry to $S^2/Z_2$ (as in projective space).\\

2D topological charges in 3D (Abrikosov vortex, fluxon) seem related with magnetic field lines (also resembling spin), there is experimentally confirmed analogy between magnetic field and vorticity~(\cite{berry,berry1,c4}). In contrast, 3D topological charge in 3D is nearly point-like and in liquid crystals get long-range interactions due to nontrivial vacuum dynamics of director field~(\cite{lq1,lq2,lq3,lq4}). We would like to propose Lagrangian recreating standard electromagnetism for them.

To calculate winding number of $\vec{n}:\mathcal{S}\to S^{2}$ in 3D by integration, we need to calculate the Jacobian. Let $u\perp v$ be unitary vectors in some point of surface $\mathcal{S}\subset \mathbb{R}^3$, transformed to $\vec{n}_u=\partial_u \vec{n}$, $\vec{n}_v=\partial_n \vec{n}$ perpendicular to $\vec{n}$, so the Jacobian is:
\be\det[\vec{n},\vec{n}_\mu,\vec{n}_\nu]=\vec{n}\cdot (\vec{n}_\mu\times \vec{n}_\nu) =\pm \|(\vec{n}\times \vec{n}_\mu)\times(\vec{n}\times \vec{n}_\nu)\|\ee
allowing to calculate \textbf{3D topological charge} like in Gauss law:
\be Q_{el}(\mathcal{S})=\frac{e_0}{4\pi} \oint_{\mathcal{S}(u,v)}du\,dv\ (\partial_u \vec{n} \times \partial_v \vec{n})\cdot \vec{n} \ee
Defining $\vec{\Gamma}_\mu=\vec{n}\times \vec{n}_\mu$ affine connection, it can be imagined as axis of local rotation for $\mu$ direction transport, with length determining its speed. Then the Jacobian becomes the curvature:
\be \vec{\Gamma}_\mu =\vec{n}\times \partial_\mu \vec{n}\qquad\qquad \vec{R}_{\mu\nu}=\vec{\Gamma}_\mu \times \vec{\Gamma}_\nu\label{con}\ee
Hence, to make  $Q_{el}(\mathcal{S})=\oint_{\mathcal{S}}E\cdot dA$ Gauss law count topological charge, we need to define electric field $E$ as curvature. To include magnetic field $B$, Faber~(\cite{faber1,faber15,faber2,faber3}) suggests to define dual (*) EM tensor with these curvatures (choose $c=1$):
\be {}^* \vec{F}_{\mu\nu}=\frac{-e_0}{4\pi \epsilon_0 c}\vec{R}_{\mu\nu}\sim
\left(
  \begin{array}{cccc}
    0 & B_1 & B_2 & B_3 \\
    -B_1 & 0 & E_3 & -E_2 \\
    -B_2 & -E_3 & 0 & E_1 \\
    -B_3 & E_2 & -E_1 & 0 \\
  \end{array}
\right)
\ee
where dual means exchanging magnetic and electric field in standard $F$ tensor: (norms of) space-space curvature corresponds to electric field, space-time to magnetic (like in vorticity-magnetic field analogy~\cite{berry,berry1}). Using standard EM Hamiltonian $\mathcal{H}_{ED} \propto \sum_{\mu\nu=0}^3 \| \vec{R}_{\mu\nu} \|^2$ leads to electromagnetism, Maxwell equations for such topological charges. For regularization there is added Higgs-like potential preferring unitary vectors for $\vec{n}$, also allowing e.g. for $\vec{n}=0$ in the center of such topological singularity e.g. hedgehog-like configuration. This field deformation to finite energy leads to Coulomb deformation in agreement with the running coupling effect~\cite{faber4}.
\subsection{Curvature for field of rotations: orthogonal matrices}
Wanting to generalize the above vector field curvature to rotations of more complex objects, let us start with describing it for orthogonal rotation matrices: $O\equiv O(x)$ field satisfying $OO^T=O^TO=I$. Transporting $\epsilon$ size step in $\mu$ direction, in linear term we get affine connection describing local rotation:
\be O\to O(I+\epsilon \Gamma_\mu)\quad\textrm{for}\quad \Gamma_\mu=O^T\, \partial_\mu O\equiv O^T O_\mu \ee
which is now anti-symmetric matrix $\Gamma_\mu=-\Gamma_\mu^T$ (from\newline $0=\partial_\mu I=\partial_\mu (O^T O)= O^T_\mu O +O^T O_\mu $).

For $3\times 3$ matrices in space, or $4\times 4$ with added 0-th coordinate as time, let us use standard notation for anti-symmetric matrix (SO(4) generator later (\ref{gb}) replaced with SO(1,3) generator by using only positive $\vec{\Gamma}^{g}_{\mu}$ for boosts):
\be\Gamma_\mu=O^T O_\mu =\left(
\begin{array}{>{\columncolor{gray!20}}cccc} \rowcolor{gray!20}
0 & \vec{\Gamma}^{g1}_{\mu}& \vec{\Gamma}^{g2}_{\mu}& \vec{\Gamma}^{g3}_{\mu}\\
-\vec{\Gamma}^{g1}_{\mu}& 0 & -\vec{\Gamma}^3_{\mu}& \vec{\Gamma}^2_{\mu} \\
-\vec{\Gamma}^{g2}_{\mu} & \vec{\Gamma}^3_{\mu} & 0 & -\vec{\Gamma}^1_{\mu}\\
-\vec{\Gamma}^{g3}_{\mu} & -\vec{\Gamma}^2_{\mu} & \vec{\Gamma}^1_{\mu} & 0 \\
\end{array}
\right) \label{g1}\ee
for the two rotation vectors built of $\Gamma_\mu=O^T O_\mu$ coordinates:
$$\vec{\Gamma}_\mu := \left((\Gamma_{\mu})_{32},(\Gamma_{\mu})_{13},(\Gamma_{\mu})_{21}\right)$$
\be \vec{\Gamma}^g_\mu := \left((\Gamma_{\mu})_{01},(\Gamma_{\mu})_{02},(\Gamma_{\mu})_{03}\right)\ee
Analogously to Faber, we would like to define dual $F$ EM tensor as proportional to  $\vec{R}_{\mu\nu}=\vec{\Gamma}_\mu \times \vec{\Gamma}_\nu$, and for GEM analogously using $\vec{R}^{gg}_{\mu\nu}=\vec{\Gamma}^g_\mu \times \vec{\Gamma}^g_\nu$ as curvature of space: submanifold perpendicular to this 0-th axis. There are also curvatures between them corresponding to EM-GEM interaction, finally we have various types of curvatures here:
\be \vec{R}_{\mu\nu}\equiv \vec{R}^{ee}_{\mu\nu}=\vec{\Gamma}_\mu \times \vec{\Gamma}_\nu\qquad
\vec{R}^{gg}_{\mu\nu}=\vec{\Gamma}^g_\mu \times \vec{\Gamma}^g_\nu \ee
$$ \vec{R}^{eg}_{\mu\nu}=\vec{\Gamma}_\mu \times \vec{\Gamma}^g_\nu\qquad\qquad \vec{R}^{ge}_{\mu\nu}=\vec{\Gamma}^g_\mu \times \vec{\Gamma}_\nu = -\vec{R}^{eg}_{\nu\mu} $$
Commutator $[\Gamma_\mu,\Gamma_\nu]=\Gamma_\mu\Gamma_\nu-\Gamma_\nu \Gamma_\mu$ of such connection matrices can be expressed with these curvatures:
\be[\Gamma_\mu,\Gamma_\nu]=\left(
\begin{array}{>{\columncolor{gray!20}}cc} \rowcolor{gray!20}
0 & -\vec{R}^{eg}_{\mu\nu}+\vec{R}^{eg}_{\nu\mu}\\
\vec{R}^{eg}_{\mu\nu}-\vec{R}^{eg}_{\nu\mu}   & \vec{R}^{ee}_{\mu\nu}+\vec{R}^{gg}_{\mu\nu} \\
\end{array}
\right):= \label{g2}\ee
\begin{scriptsize}
$$\left(
\begin{array}{>{\columncolor{gray!20}}cccc} \rowcolor{gray!20}
0 & -\vec{R}^{eg1}_{\mu\nu}+\vec{R}^{eg1}_{\nu\mu}& -\vec{R}^{eg2}_{\mu\nu}+\vec{R}^{eg2}_{\nu\mu}& -\vec{R}^{eg3}_{\mu\nu}+\vec{R}^{eg3}_{\nu\mu}\\
\vec{R}^{eg1}_{\mu\nu}-\vec{R}^{eg1}_{\nu\mu}   & 0 & -\vec{R}^{ee3}_{\mu\nu}-\vec{R}^{gg3}_{\mu\nu}& \vec{R}^{ee2}_{\mu\nu}+\vec{R}^{gg2}_{\mu\nu} \\
\vec{R}^{eg2}_{\mu\nu}-\vec{R}^{eg2}_{\nu\mu} & \vec{R}^{ee3}_{\mu\nu}+\vec{R}^{gg3}_{\mu\nu} & 0 & -\vec{R}^{ee1}_{\mu\nu}-\vec{R}^{gg1}_{\mu\nu}\\
\vec{R}^{eg3}_{\mu\nu}-\vec{R}^{eg3}_{\nu\mu} & -\vec{R}^{ee2}_{\mu\nu}-\vec{R}^{gg2}_{\mu\nu} & \vec{R}^{ee1}_{\mu\nu}+\vec{R}^{gg1}_{\mu\nu} & 0 \\
\end{array}
\right) $$
\end{scriptsize}

\noindent with shortened first matrix notation, expanded in the latter.

In flat spacetime we can express this commutator as:
$$0=\partial_\mu \partial_\nu O -\partial_\nu \partial_\mu O=\partial_\mu (O\Gamma_\nu)-\partial_\nu (O\Gamma_\mu)$$
\be [\Gamma_\mu,\Gamma_\nu] = \partial_\nu \Gamma_\mu - \partial_\mu \Gamma_\nu\ee

We could use $[\Gamma_\mu,\Gamma_\nu]$ as $F$ tensor in Lagrangian, but it seems missing crucial factors required e.g. to make hedgehog in space (like electron) much lighter than in time (like black hole) - added next by lengths of axes as in Landau-de Gennes model.

\section{Extension to bixial nematic as ellipsoid field}\label{sec3}
Vector field does not recognize twists of such vectors, hence would allow violation of angular momentum in this direction. In liquid crystals it is repaired by going from (uniaxial) Oseen-Frank model seen as approximation, to full (biaxial) Landau-de Gennes model~\cite{gennes} - field of ellipsoids of potentially three different axes, offering simple general description of field recognizing all SO(3) rotations (also allowing further search "of what" for even more fundamental models), we will further expand with boosts to SO(1,3).

Like e.g. stress-energy tensor, such objects can be represented using real symmetric matrix/tensor field $M\equiv M(x)$ with chosen shape as preferred set of eigenvalues representing lengths of the 3 axes, its eigenvectors point directions of these 3 axes:
\be M=ODO^T\qquad\textrm{for } OO^T=I,\ D=\textrm{diag}(\lambda_1,\lambda_2,\lambda_3)\ee
for $O$ orthogonal matrix field as in the previous subsection. We will focus on 3D case now, but it naturally generalizes to 4D case using $4\times 4$ matrices, $D=\textrm{diag}(\lambda_0,\lambda_1,\lambda_2,\lambda_3)$, and $O$ from SO(1,3) containing boost (no longer orthogonal).

The diagonal matrix $D$ should prefer some shape: e.g. fixed $(\Lambda_0\geq)\ \Lambda_1 \geq \Lambda_2 \geq\Lambda_3$. However, it also requires a possibility of regularization of singularities to finite energy (like top of Fig. \ref{intr} in 2D), what again can be obtained using Higgs-like potential, this time with SO(3) minimum, for example:
\be V(M)=\sum_i (\lambda_i-\Lambda_i)^2\quad\textrm{or e.g. }\quad\sum_{k=1}^3 (\textrm{Tr}(M^k)-c_k)^2\ee
for $c_k=\sum_i (\Lambda_i)^k$ like in Landau-de Gennes potential~\cite{gennes}: \be V_{LG}(M)= a\,\textrm{Tr}(M^2)-b\,\textrm{Tr}(M^3)+c\,\left(\textrm{Tr}(M^2)\right)^2\ee This potential is supposed to be activated mainly near particles to prevent infinity (regularization) - corresponds to weak/strong interaction, hence the choice of its details remains a difficult open question requiring simulations.

Let us now focus  on the $D\approx \textrm{diag}(\Lambda_1,\Lambda_2,\Lambda_3)$ vacuum behavior. Thermally it should also contain tiny perturbations, which  might correspond to dark energy/matter in analogy to 2.7K cosmic microwave background radiation.

Derivative in $\mu$ direction of our tensor field is:
\be M_\mu := \partial_\mu M = O_\mu D O^T + ODO^T_\mu + OD_\mu O^T \label{dM} \ee
$$ O^T M_\mu O = O^T O_\mu D +D O^T_\mu O + D_\mu = \Gamma_\mu D - D\Gamma_\mu+D_\mu$$
for $\Gamma_\mu=O^T O_\mu$ affine connection being anti-symmetric matrix as in the previous subsection.

From dynamics of rotation part $O$ (later in 4D including boosts), as in Faber model we would like to define EM field as its curvature to make Gauss law count winding number. However, for a few reasons like distinguishing rotations of various energy (EM $\gg$ pilot wave $\gg$ GEM). Therefore, instead of $O$ as previously, this time we would like to directly work on $M(x)\equiv M=ODO^T$ field.
\subsection{Curvature analogue of electromagnetic $F$ tensor}
While there might be a better choice, for now let us focus on a simple one: try to just replace discussed previously $[\Gamma_\mu,\Gamma_\nu]$ commutator with $[M_\mu,M_\nu]=M_\mu M_\nu - M_\nu M_\mu$:
\be F_{\mu\nu}:=[M_\mu,M_\nu]\quad =-F_{\nu\mu}=-F_{\mu\nu}^T\ee
for $M_\mu :=\partial_\mu M$. Looking at (\ref{dM}), focusing on vacuum dynamics $D=\textrm{diag}(\Lambda_1,\Lambda_2,\Lambda_3)$ and conveniently transforming:
\begingroup
\renewcommand*{\arraystretch}{1.5}
\be O^T F_{\mu\nu} O=O^T[M_\mu, M_\nu]O\approx [\Gamma_\mu D - D\Gamma_\mu,\Gamma_\nu D - D\Gamma_\nu]=\label{RF}\ee
$$(\Lambda_1-\Lambda_2)(\Lambda_3-\Lambda_1)(\Lambda_2-\Lambda_3)
\begin{pmatrix}
0 & \frac{-\vec{R}^3_{\mu\nu}}{\Lambda_1-\Lambda_2}& \frac{\vec{R}^2_{\mu\nu}}{\Lambda_3-\Lambda_1}\\                                                                            \frac{\vec{R}^3_{\mu\nu}}{\Lambda_1-\Lambda_2}& 0 & \frac{-\vec{R}^1_{\mu\nu}}{\Lambda_2-\Lambda_3}\\                                                                            \frac{-\vec{R}^2_{\mu\nu}}{\Lambda_3-\Lambda_1} & \frac{\vec{R}^1_{\mu\nu}}{\Lambda_2-\Lambda_3} & 0 \\                                                                         \end{pmatrix}                                                                     $$
\endgroup
for $\vec{R}_{\mu\nu}=\vec{\Gamma}_\mu \times \vec{\Gamma}_\nu$ and $\vec{\Gamma}_\mu := \left((\Gamma_{\mu})_{32},(\Gamma_{\mu})_{13},(\Gamma_{\mu})_{21}\right)$ as in the previous subsection. Calculating topological charge through integration using such matrix curvature, we get charge quantization for its each coordinate.

Wanting to interpret $F_{\mu\nu}$ EM tensor with such curvature, instead of single number (or vector in Faber approach), it is now anti-symmetric matrix (or SO(1,3) generator in 4D), requiring to replace $\|F_{\mu\nu}\|$ with a matrix norm. A natural generalization is Frobenius inner product and norm, treating matrix as vector for Euclidean norm:

\be A\bullet B=\textrm{Tr}(A B^T)=\sum_{ij} A_{ij} B_{ij}\qquad \|A\|_F=\sqrt{A\bullet A}\ee

Prolate uniaxial nematic director field case can be imagined as $\Lambda_1> \Lambda_2=\Lambda_3$ limit, leaving single curvature $\|R_{\mu\nu}\|_F^2 \propto (\vec{R}_{\mu\nu,1})^2$, making $R_{\mu\nu}^1\equiv \vec{R}_{\mu\nu,1}$ proportional to electric field for spatial $1\leq \mu < \nu < 3$, and to magnetic field for temporal $\mu=0$ and spatial $\nu\in \{1,2,3\}$.

For promising similar biaxial nematic case $\Lambda_1> \Lambda_2>\Lambda_3$, we would like $\Lambda_2\approx \Lambda_3$ being much closer as in Fig. \ref{intr},\ref{diagram} - adding to electromagnetism ($R^1$ with dominant first axis), low energy degrees of freedom: $R_{\mu\nu}^2\equiv \vec{R}_{\mu\nu,2}, R_{\mu\nu}^3\equiv \vec{R}_{\mu\nu,3}$, hopefully to agree with quantum phase for pilot wave, propelled by particle configuration (de Broglie clock).
\subsection{Lagrangian, four-potential, uniaxial as special case}
Let us postulate the \textbf{Lagrangian} in analogy to EM:
\be \mathcal{L}=\sum_{\mu=1}^3 \|F_{\mu 0}\|_F^2 - \sum_{1\leq \mu<\nu\leq 3} \|F_{\mu\nu}\|_F^2 -V(M)  \label{lagrangian}\ee
\subsubsection{\textbf{Four-potential $A_\mu$} analogue} The mentioned suggestion to directly use $F_{\mu\nu}=[M_\mu,M_\nu]$ in Lagrangian is inconvenient due to products of derivatives. Hence let us introduce $A_\mu$ to work as EM four-potential, this time being matrices ($3\times 3$ or $4\times 4$ with gravity). $A_\mu = M\, \partial_\mu M$ would already give $\partial_\mu A_\nu - \partial_\nu A_\mu=[M_\mu,M_\nu]$, but let us anti-symmetrize it for reduced dimension and direct interpretation:
\be F_{\mu\nu}=\partial_\mu A_\nu - \partial_\nu A_\mu\quad\textrm{for}\quad A_\mu := M M_{\mu}-M_\mu M \approx \label{Amat}\ee
\begin{scriptsize}
$$\approx O
\begin{pmatrix}
0 & \vec{\Gamma}^3_{\mu}(\Lambda_1-\Lambda_2)^2 & -\vec{\Gamma}^2_{\mu}(\Lambda_1-\Lambda_3)^2\\                                                                            -\vec{\Gamma}^3_{\mu}(\Lambda_1-\Lambda_2)^2& 0 & \vec{\Gamma}^1_{\mu}(\Lambda_2-\Lambda_3)^2\\                                                                            \vec{\Gamma}^2_{\mu}(\Lambda_1-\Lambda_3)^2 & -\vec{\Gamma}^1_{\mu}(\Lambda_2-\Lambda_3)^2& 0 \\                                                                         \end{pmatrix}
O^T$$
\end{scriptsize}
\noindent where the $A_\mu$ approximation is again for  $D=\textrm{diag}(\Lambda_1,\Lambda_2,\Lambda_3)$ vacuum situation. Using commutation of derivatives we get analogue of Maurer-Cartan structural equation:
\be \partial_\mu A_\nu - \partial_\nu A_\mu=2(\partial_\mu M\, \partial_\nu M- \partial_\nu M\, \partial_\mu M)=:2 F_{\mu\nu} \ee
We can calculate variation, which is real anti-symmetric matrix:
\be \frac{\partial \|F_{\mu\nu}\|_F^2}{\partial(\partial_\alpha A_\beta)}=\frac{1}{2}\frac{\partial \|\partial_\mu A_\nu - \partial_\nu A_\mu\|_F^2}{\partial(\partial_\alpha A_\beta)}=(\delta_{\mu\alpha}\delta_{\nu\beta}+\delta_{\mu\beta}\delta_{\nu\alpha})F_{\alpha\beta}\label{ad}\ee
\subsubsection{\textbf{Uniaxial nematic} as degenerate case} Director $\vec{n}$ field can be obtained using e.g. $M=\vec{n}\,\vec{n}^T$ corresponding to $\Lambda_1 =\|\vec{n}\|^2$ (constant), $\Lambda_2=\Lambda_3=0$ case:
$$A_\mu= \|\vec{n}\|^2 \begin{pmatrix}
0 & (\vec{n}\times \vec{n}_\mu)_3 & -(\vec{n}\times \vec{n}_\mu)_2\\                                                                            -(\vec{n}\times \vec{n}_\mu)_3 & 0 & (\vec{n}\times \vec{n}_\mu)_1\\                                                                            (\vec{n}\times \vec{n}_\mu)_2 & -(\vec{n}\times \vec{n}_\mu)_3& 0 \\                                                                         \end{pmatrix}$$
For which $\partial_\mu A_\nu - \partial_\nu A_\mu$ gives curvature as in Faber approach:
$$\partial_\mu (\vec{n}\times \vec{n}_\nu)-\partial_\nu (\vec{n}\times \vec{n}_\mu)=2\,\vec{n}_\mu \times \vec{n}_\nu$$

Here we have 2 vacuum degrees of freedom rotating $\vec{n}$, in general case we slightly separate $\Lambda_2$ and $\Lambda_3$ by adding one low energy degree of freedom for twists of $\vec{n}$, supposed to work as quantum phase. To see this generalization as perturbation, we will replace $\Lambda=(1,0,0)$ case with $\Lambda=(1,\delta,0)$ for tiny $\delta$ related with Planck constant, and focus on low order $\delta$ terms.
\section{General equations of motion for 3D case} \label{sec4}
Let us now derive equations of motion from Lagrangian optimization - zeroing of its variation. For EM it is usually done with variation of $A$ field - we will start with as simpler. However, e.g. to get built-in charge quantization, there was proposed more fundamental field $M$ (which topological charge is calculated by Gauss law) - we further consider its variation.
\subsection{Simplification: Euler-Lagrange equations for $A$ field}
Equations of motion for electromagnetism are usually derived with Euler-Lagrange equations for $A$ field (\ref{ad}) - let us start here as simpler, bringing valuable intuitions:
$$\frac{\partial \mathcal{L}}{\partial A_\alpha}=\frac{d}{dx_0} \frac{\partial\mathcal{L}}{\partial(\partial_0 A_{\alpha})}+\sum_{i=1}^3 \frac{d}{dx_i} \frac{\partial\mathcal{L}}{\partial(\partial_i A_{\alpha})}$$
\be\frac{\partial V}{\partial A_\alpha}= \partial_0 F_{0 \alpha } - \sum_{i=1}^3 \partial_i F_{i \alpha}=\Box A_\alpha -\partial_\alpha \left(\partial_0 A_0-\sum_i \partial_i A_i\right) \ee
\noindent for $\Box = \partial_{00}-\sum_{i=1}^3 \partial_{ii}$ d'Alembertian.

In vacuum the potential vanishes, getting Maxwell-like equations for $\vec{E}=(F_{23},F_{31},F_{12})$, $\vec{B}=(F_{01},F_{02},F_{03})$ dual analogs of electric and magnetic fields, but this time with each component being a matrix, in vacuum satisfying $\frac{\partial V}{\partial A_\alpha}=0= \Box F_{\mu\nu}$ wave equation with $c=1$ propagation speed, and with built-in charge quantization as topological.

As in EM Lorentz gauge condition, the $\partial_0 A_0-\sum_i \partial_i A_i = [M,\Box M]$ term should be zero from integration by parts (assuming fields vanish in infinity), leading to $\frac{\partial V}{\partial A_\alpha}=\Box A_\alpha$.

Regarding the \textbf{potential}, its choice remains difficult main open question, which will require simulations e.g. aiming agreement with electron, 3 leptons.

While there was mentioned potential $V(M)$ directly preferring shape as $(\lambda_i)$ (similarity to Faber), here we get $\frac{\partial \mathcal{L}}{\partial A_\alpha}$ suggesting to use $V(A)$ instead - as in (\ref{Amat}) using differences of $(\lambda_i)$, this time multiplied by derivative in $\Gamma$.

Both choices have $M$ derivative dependence in $\Gamma$, which if preferring some values with Higgs-like potential, enforce nonzero $M$ derivatives - what might be the source e.g. of Zitterbewegung intrinsic periodic process of electron~\cite{clock}.

Ideally would be not having to fix shape $(\Lambda_i)$ as parameters of the model, but to make them automatically emerge from a simple e.g. $V(A)=(\sum_\mu \|A_\mu\|_F^2 -1)^2$ Higgs-like potential, with additional e.g. volume constraint $\det(M)=\prod_i \lambda_i = \textrm{const}$ to prevent using only long axes which allow for low curvature (hence energy).

\textbf{Hamiltonian} (energy density) derivation is analogous to EM:
$$\mathcal{H}=\sum_{\mu=1}^3 \frac{\partial \mathcal{L}}{\partial(\partial_0 A_\mu)}\partial_0 A_\mu-\mathcal{L}=\sum_{\mu=1}^3 F_{0\mu}\bullet (2F_{0\mu}+\partial_\mu A_0)-\mathcal{L}$$
\be\mathcal{H}=\sum_{0\leq \mu<\nu\leq 3} \|F_{\mu\nu}\|_F^2 +V\quad+
\sum_{\mu=1}^3 F_{0\mu}\bullet \partial_\mu A_0 \label{hamil}\ee
The last sum vanishes in EM due to integration by parts (assuming fields vanish in infinity) to shift derivative to $F$, getting divergence of electric field without sources. Here it becomes $-A_0\bullet \sum_{\mu=1}^3 \partial_\mu F_{0\mu}$, which from above Euler-Lagrange equation vanishes at least in vacuum.

\subsection{Proper equations of motion: variation of $M$ field}
In the discussed approach, as in Fig. \ref{diagram}, we assume there is more fundamental field $M$ - e.g. to make Gauss law count its topological charge, enforcing charge quantization.

To get equations of motion we consider its variation. In 3D we have 3 rotation generators $G$:
\be\left( \begin{array}{ccc}
                  0 & 0 & 0 \\   0 & 0 & -1 \\   0 & 1 & 0 \\
                \end{array} \right),
\left(    \begin{array}{ccc}
                  0 & 0 & 1 \\     0 & 0 & 0 \\  -1 & 0 & 0 \\
                \end{array}    \right),
                \left(   \begin{array}{ccc}
                  0 & -1 & 0 \\   1 & 0 & 0 \\   0 & 0 & 0 \\
                \end{array}    \right)\label{rotgen}\ee
(plus 3 in 4D), and 3 (+1 in 4D) axis elongation generators:
\be \left( \begin{array}{ccc}
                  1 & 0 & 0 \\   0 & 0 & 0 \\     0 & 0 & 0 \\
                \end{array}  \right), \left(  \begin{array}{ccc}
                  0 & 0 & 0 \\    0 & 1 & 0 \\    0 & 0 & 0 \\
                \end{array}   \right),\left(    \begin{array}{ccc}
                  0 & 0 & 0 \\  0 & 0 & 0 \\     0 & 0 & 1 \\
                \end{array}   \right) \label{scale} \ee
Now for $M=ODO^T$ field, generator as matrix $G$ (one of 3+3 above, 3+3+4 in 4D), infinitesimal $\epsilon \in \mathbb{R}$, $\eta:\mathbb{R}^4\to \mathbb{R}$ function, let us consider variation in convenient form where generators are directly acting on $D$:
\be O \to O(I+\epsilon \eta G)\label{varia}\ee
$$M\to O(I+\epsilon \eta G)\,D\, (I+\epsilon \eta G^T)O^T\approx M+\epsilon \eta O(GD+DG^T)O^T $$
\be M \to M+ \epsilon\eta O G' O^T \qquad \textrm{for}\qquad G'=GD+DG^T \ee
\be M_\mu\to M_\mu +\epsilon O(\eta_\mu G'+\eta [\Gamma_\mu,G']+\eta G'_\mu)O^T \label{Mmuv}\ee
plus $\textsf{O}(\epsilon^2)$ neglected higher order term.


We will work on $G'=GD+DG^T$ 6 (or 10 in 4D) generators, which for rotations have two $+1$ coefficients, for elongations we can take $G'=G$.\\

\subsubsection{Vacuum case derivation - fixed $D$, only SO(3) rotations}
For simplicity let us focus first on vacuum case: fixed $\lambda_i =\Lambda_i$ minimizing potential, only 3 rotation generators (3+3 in 4D). We also use $G'_\mu=0$, but nonzero is included in the final formula (\ref{feq}).

Using $\Gamma_\mu =O^T O_\mu=-O^T_\mu O$, for rotations only we can get conveniently transformed versions: $\overline{M}_\mu, \overline{A}_\mu, \overline{F}_{\mu\nu}$ with commutators:
\be M_\mu =\partial_\mu (ODO^T)= O \overline{M}_\mu O^T \quad \textrm{for}\quad \overline{M}_\mu=[\Gamma_\mu,D]\ee
\be A_\mu=[M,M_\mu]=O  \overline{A}_\mu O^T \quad\textrm{for}\quad \overline{A}_\mu=[D,\overline{M}_\mu] \ee
\be F_{\mu\nu} =[M_\mu,M_\nu]=O  \overline{F}_{\mu\nu} O^T \quad\textrm{for}\quad \overline{F}_{\mu\nu} =[\overline{M}_\mu,\overline{M}_\nu]\label{rotM}\ee

Applying variation (\ref{varia}) and denoting $G':=[G,D]$:
\be M \to M+\epsilon \eta\, O [G,D] O^T =O(D+\epsilon \eta G')O^T \label{mm}\ee
plus  $\textsf{O}(\epsilon^2)$. As $\overline{M}_\mu:=[\Gamma_\mu,D]$, its $\partial_\mu$ derivative is:
\be \partial_\mu M =O \overline{M}_\mu O^T \to O\left(\overline{M}_\mu  +\epsilon\eta[\Gamma_\mu,G']+\epsilon \eta_\mu G' \right)O^T\ee
plus $\textsf{O}(\epsilon^2)$. Using $F_{\mu\nu}=[M_\mu,M_\nu]$, $\overline{F}_{\mu\nu}=O^T F_{\mu\nu} O$:
$$\overline{F}_{\mu\nu}\to \left[\,\overline{M}_\mu  +\epsilon\eta[\Gamma_\mu,G']+\epsilon \eta_\mu G',\ \overline{M}_\nu  +\epsilon\eta[\Gamma_\nu,G']+\epsilon \eta_\nu G'\right]$$
$$=\overline{F}_{\mu\nu}  +\epsilon\eta([\overline{M}_\mu,[\Gamma_\nu,G']]-[\overline{M}_\nu,[\Gamma_\mu,G']])+ $$
$$+\epsilon\eta_\nu [\overline{M}_\mu,G']-\epsilon\eta_\mu[\overline{M}_\nu,G']+\textsf{O}(\epsilon^2)$$
Lagrangian (\ref{lagrangian}) needs $-\textrm{Tr}(F_{\mu \nu} F^T_{\mu\nu})=\textrm{Tr}(F_{\mu \nu} F_{\mu\nu})\to$
$$\textrm{Tr}(F_{\mu \nu} F_{\mu\nu}) +2\epsilon \eta \textrm{Tr}\left(\overline{F}_{\mu\nu}\left( [\overline{M}_\mu,[\Gamma_\nu, G']] -[\overline{M}_\nu,[\Gamma_\mu, G']] \right)\right)+    $$
$$+2\epsilon  \textrm{Tr}\left(\eta_\nu\overline{F}_{\mu\nu} [\overline{M}_\mu,G'] - \eta_\mu \overline{F}_{\mu\nu} [\overline{M}_\nu,G']\right)+\textsf{O}(\epsilon^2)$$
using $\textrm{Tr}(F_{\mu \nu} F_{\mu\nu})=\textrm{Tr}(\overline{F}_{\mu \nu} \overline{F}_{\mu\nu})$, $\textrm{Tr}(AB)=\textrm{Tr}(BA)$.
Lagrangian (\ref{lagrangian}) sums 6 $\textrm{Tr}(F_{\mu \nu} F_{\mu\nu})$ terms. As in the minimum necessary condition, we get the least action if the $\epsilon$ term vanishes. Like in derivation of Euler-Lagrange equation, we need first to apply integration by parts to shift $\eta$ derivatives (assuming $\eta$ vanishes at some boundary).  Using $ \partial_\mu M_\nu =\partial_\nu M_\mu$:
$$ \partial_\nu \overline{M}_\mu = \partial_\nu \left(O^T M_\mu O\right) =O_\nu^T M_\mu O+O^T (\partial_\nu M_{\mu}) O+O^T M_\mu O_\nu  $$
$$=\Gamma_\nu^T \overline{M}_\mu  + O^T (\partial_\nu M_{\mu}) O +\overline{M}_{\mu} \Gamma_\nu = O^T (\partial_\nu M_{\mu}) O + [\overline{M}_{\mu}, \Gamma_\nu]$$
\be \partial_\nu \overline{M}_\mu - \partial_\mu \overline{M}_\nu=[\overline{M}_{\mu}, \Gamma_\nu]-[\overline{M}_{\nu}, \Gamma_\mu]\ee
Finally the $\delta\mathcal{L}=0$ equations of motion are:
$$ 0=\sum_{\mu\nu} d_{\mu\nu} \textrm{Tr}\left(\overline{F}_{\mu\nu}\left( [\overline{M}_\mu,[\Gamma_\nu, G']] -[\overline{M}_\nu,[\Gamma_\mu, G']]\right)\right)-$$
$$-\textrm{Tr}\left(\overline{F}_{\mu\nu} \left([[\overline{M}_{\mu}, \Gamma_\nu], G']-[[\overline{M}_{\nu}, \Gamma_\mu],G']\right)\right)) $$
$$ -\textrm{Tr}\left(\overline{F}_{\mu\nu,\mu} [\overline{M}_\nu,G'] - \overline{F}_{\mu\nu,\nu} [\overline{M}_\mu,G']\right)$$
for $\overline{F}_{\mu\nu,\mu}=\partial_\nu \overline{F}_{\mu\nu}$, $d_{\mu\nu}=1$ for $\mu\nu \in \{10,20,30\}$ and  $d_{\mu\nu}=-1$ for $\mu\nu \in \{23,31,12\}$ and 0 otherwise. Using Jacobi identity $([[A,B],C]+[[B,C],A]+[[C,A],B]=0)$ we can simplify the first two lines:
\be 0=\sum_{\mu\nu} d_{\mu\nu} \textrm{Tr}\left(\overline{F}_{\mu\nu}\left( [\Gamma_\nu,[\overline{M}_\mu,G']] -[\Gamma_\mu,[\overline{M}_\nu,G']]\right)\right)+\label{feq}\ee
$$ + \textrm{Tr}\left(\overline{F}_{\mu\nu,\nu} [\overline{M}_\mu,G']-
\overline{F}_{\mu\nu,\mu} [\overline{M}_\nu,G']
\right)$$
In the general case there there is additional $G'_\mu =\partial_\mu G' =GD_\mu+D_\mu G$ term in (\ref{Mmuv}) depending on evolution of diagonal $\partial_\mu D = D_\mu$. It needs additional $\overline{F}_{\mu\nu} ([M_\mu,G'_\nu]-[M_\nu,G'_\mu])$ term in (\ref{feq}), and including potential $V$.
\begin{figure*}[t!]
    \centering
        \includegraphics[width=18.5cm]{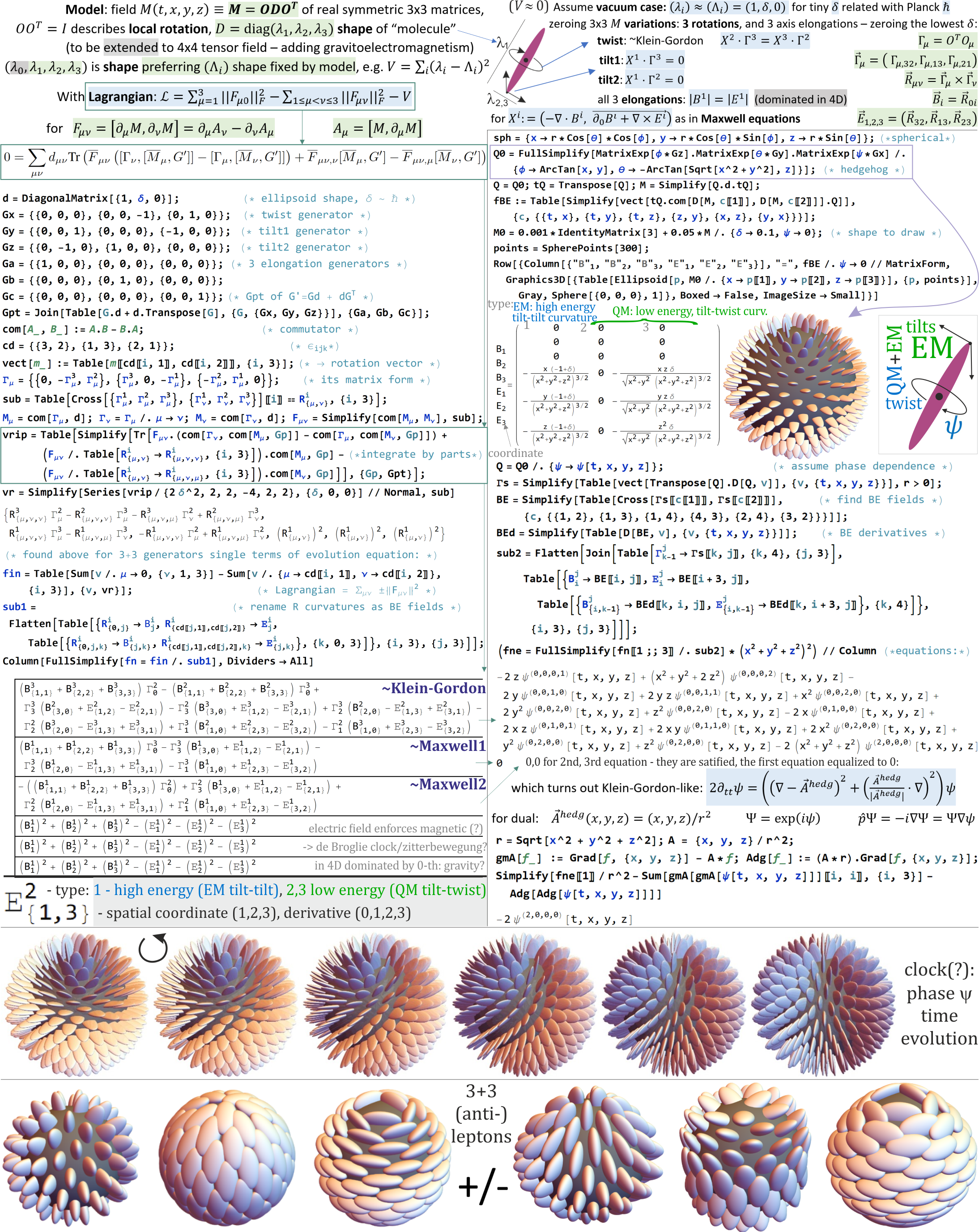}
        \caption{Used Mathematica source for 3D case and $\Lambda =(1,\delta,0)$ shape (available in \url{https://github.com/JarekDuda/liquid-crystals-particle-models}). Application of written evolution equation (\ref{feq}) $M_\mu$ in vacuum case ($V=0,G'_\mu=0$), 6 generators $G$, and  neglecting higher $\delta$ terms. Right: application for basic hedgehog case with $\psi$ function defining local twist (wavefunction $\Psi=\exp(i\psi)$), deriving Klein-Gordon-like equation. Bottom: visualization of expected phase evolution for de Broglie clock/Zitterbewegung, and of 3 leptons, anti-leptons. }
        \label{source}
\end{figure*}

\subsubsection{Simplification}:
These equations for 3 generators $G$ are still quite complex. To simplify as in (\ref{RF}), denote 3x3 anti-symmetric matrices using $\vec{R}_{\mu\nu}=\vec{\Gamma}_\mu \times \vec{\Gamma}_\nu $ vectors. Then denote EB fields as coordinates of $R=F^*$ tensor:
\be\vec{B}_1=\vec{R}_{01}\qquad \vec{B}_2=\vec{R}_{02}\qquad\vec{B}_3=\vec{R}_{03}\label{BEfields}\ee
$$\vec{E}_1=\vec{R}_{32}\qquad \vec{E}_2=\vec{R}_{13}\qquad\vec{E}_3=\vec{R}_{21}$$
each of them is now 3D vector, which coordinates correspond to different energies as in Fig. \ref{diagram} - let us denote them with superscript e.g. $\vec{B}_i=(B_i^1,B_i^2,B_i^3)$.

Let us now choose $\Lambda$ eigenvalues as $D=\textrm{diag}(1,\delta,0)$. For $\delta=0$ we get uniaxial nematic Faber's case. Here we assume $\delta$ is tiny positive, with twist corresponding to quantum phase - hence $\delta$ should be related with Planck constant. Neglecting higher order $\delta$ terms, the 3 EB coordinates correspond to $\approx (1,\delta,\delta)$ energies: first coordinate $(B^1=(B^1_1,B^1_2,B^1_3), E^1=(E^1_1,E^1_2,E^1_3))$ to standard electromagnetism, the remaining two to low energy quantum phase, hopefully to recreate relativistic QM like Klein-Gordon, QED Lagrangian.

To find the final equations (\ref{feq}), neglecting higher order $\delta$ terms, there is provided used Mathematica source (GitHub, Fig. \ref{source}), leading to terms for 3 rotation generators:
$$\textrm{twist:}\quad \Gamma_\mu^2 \partial_\nu R_{\mu\nu}^3- \Gamma_\mu^3 \partial_\nu R_{\mu\nu}^2-\Gamma_\nu^2 \partial_\mu R_{\mu\nu}^3+\Gamma_\nu^3 \partial_\mu R_{\mu\nu}^2$$
$$\textrm{2 tilts:}\quad\Gamma_\mu^3 \partial_\nu R_{\mu\nu}^1- \Gamma_\nu^3 \partial_\mu R_{\mu\nu}^1,\quad
\Gamma_\nu^2 \partial_\mu R_{\mu\nu}^1- \Gamma_\mu^2 \partial_\nu R_{\mu\nu}^1$$
Such terms need to be summated by $\mu,\nu$ and equalized to 0, leading to Maxwell-like equation terms for EM. Denoting:
\be X^i=(-\nabla\cdot B^i, \overrightarrow{\partial_0 B^i+\nabla\times E^i}),\ee
the (\ref{feq}) equations for 3 rotation generators become:
\be X^2\cdot \Gamma^3 = X^3\cdot \Gamma^2\qquad \sim\textrm{Klein-Gordon for twist(phase)}  \ee
$$ X^1 \cdot \Gamma^3 = 0,\ X^1 \cdot \Gamma^2 = 0\qquad \textrm {Maxwell equation for two tilts}$$
Figure \ref{source} contains this implementation, also applying these equations to hedgehog ansatz (model of lepton), getting Klein-Gordon-like equation for the twist (phase). Figure \ref{4Deq} contains implementation deriving 4D equations, getting 2nd set of Maxwell-like equations for GEM. Figure \ref{coulomb} calculates Coulomb effective potential for such two topological charges, Figure \ref{newton} suggests a way to analogously get Newton law for 4D field.

Further work is planned to extend this agreement, also parametrization to moduli space, finally maybe hydrodynamical simulations. The first difficulty is getting angular momentum, clock for charge (electron), hopefully through regularization by including Higgs-like potential, as most of mass/energy of particle is localized in its center - where  potential is nonzero.

\begin{figure*}[t!]
    \centering
        \includegraphics[width=18.5cm]{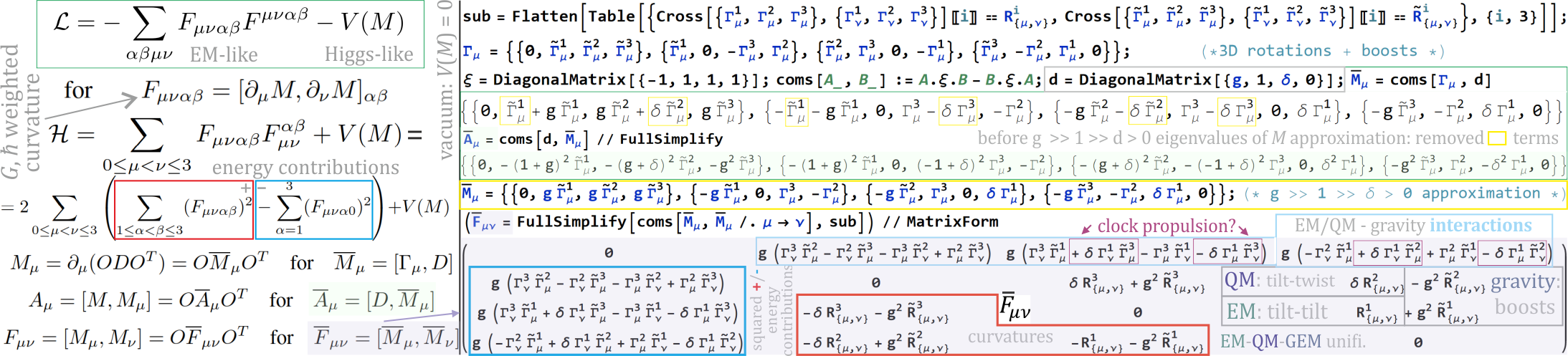}
        \caption{Vacuum ($V(M)=0$) calculation (in GitHub) for suggested 4D LdGS model, $M=ODO^T$, $D=\textrm{diag}(g,1,\delta,0)$, $g\gg 1\gg\delta>0$, $O\in$ SO(1,3) with shown approximation for $\bar{F}_{\mu\nu}$. By tilde there are marked time axis/gravitational boosts, $\vec{R}_{\mu\nu}=\vec{\Gamma}_\mu \times \vec{\Gamma}_\nu$. Surprisingly, the Hamiltonian turns out having not only positive (red), but also negative (blue) contributions. Positive energy contributions (red) for separate EM and GEM are as expected ($\sim E^2+B^2$), lead to Maxwell equations for separate each of them (Fig. \ref{source}, \ref{4Deq}). Combined lead to EM/QM-GEM interactions, like slowing down of EM propagation in GEM field for light lensing by Fermat principle, and gravitational time dilation. Additionally, the negative energy contributions (blue) $\Gamma \tilde{\Gamma}$ give tendency both for gravitational mass (local boosts in presence of particles) and de Broglie clock/angular momentum of e.g. electron ($\Gamma_0$ time derivative). Finding  $g,\delta$ is surprisingly difficult, comparing with QED Lagrangian suggests approximate $\delta^2 \sim \hbar c$, for gravity by cost of boosts: $g^4 \sim ke^2/Gm^2\approx 10^{38}$ for electron.  }
        \label{Fmunu}
\end{figure*}

\section{4D case: Lorentz invariance and gravity} \label{sec5}

\begin{figure}[t!]
    \centering
        \includegraphics[width=8.5cm]{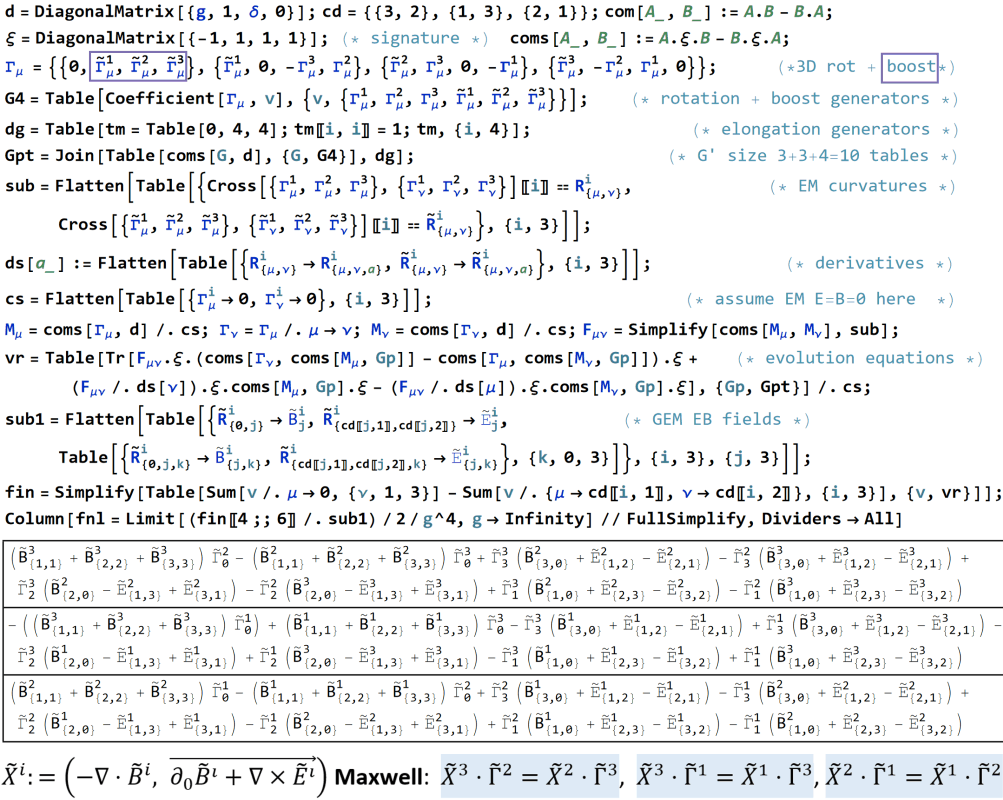}
        \caption{Basic source to derive equations for 4D case (available in GitHub) with $\Lambda =(g,1,\delta,0)$ shape and $g\gg 1$. Now we have 10 generators: 3 for 3D rotations corresponding to EM, 3 for boosts as rotations of 0th axis (symmetric vs antisymmetric generators) - leading to GEM, and 4 equations for elongations of 4 axes ($\tilde{\Gamma}\equiv \Gamma^g, \tilde{R}\equiv R^{gg}$). Such equations become much more complex, there is derived GEM-only case with zeroed 3D rotations, getting Maxwell-like equations for GEM.}
        \label{4Deq}
\end{figure}
Previously we were focused on 3D case as for biaxial nematic, briefly mentioning 4D case e.g. as SO(4) rotation in (\ref{g1}). In contrast, there is a general belief for Lorentz invariance in 4D, replacing SO(4) with SO(1,3) Lorentz group. The previous antisymmetric rotation generator (\ref{g1}) with mixed below - all positive $\vec{\Gamma}^{g}_{\mu}$ being boost generator for chosen rapidity:
\be\Gamma_\mu=O^T O_\mu =\left(
\begin{array}{>{\columncolor{gray!20}}cccc} \rowcolor{gray!20}
0 & \vec{\Gamma}^{g1}_{\mu}& \vec{\Gamma}^{g2}_{\mu}& \vec{\Gamma}^{g3}_{\mu}\\
\vec{\Gamma}^{g1}_{\mu}& 0 & -\vec{\Gamma}^3_{\mu}& \vec{\Gamma}^2_{\mu} \\
\vec{\Gamma}^{g2}_{\mu} & \vec{\Gamma}^3_{\mu} & 0 & -\vec{\Gamma}^1_{\mu}\\
\vec{\Gamma}^{g3}_{\mu} & -\vec{\Gamma}^2_{\mu} & \vec{\Gamma}^1_{\mu} & 0 \\
\end{array}
\right) \label{gb}\ee
Matrix $O\equiv O_\alpha^\beta$ contains rotation and boost, can be calculated by exponentiation of above generator matrix. It is no longer orthogonal, but still we can use $M=ODO^T$ as rotation and boost of some fundamental anisotropic object. 

In Euler-Lagrange equations for vacuum we should use 6 generators: 3 antisymmetric for 3D rotation (\ref{rotgen}), and 3 symmetric for boosts (+ 4 axis elongation generators):
$$\left( \begin{array}{cccc}
                  0 & 1 & 0&0 \\   1 & 0 &0&0 \\   0 & 0 &0&0 \\ 0 & 0 &0&0 \\
                \end{array} \right),
\left( \begin{array}{cccc}
                  0 & 0 & 1&0 \\   0 & 0 &0&0 \\   1 & 0 &0&0 \\ 0 & 0 &0&0 \\
                \end{array} \right),\left( \begin{array}{cccc}
                  0 & 0 & 0&1 \\   0 & 0 &0&0 \\   0 & 0 &0&0 \\ 1 & 0 &0&0 \\
                \end{array} \right)$$

Another change is in matrix products - they are unchanged for $O\equiv O_\alpha^\beta$ up-down indexes. However, our fundamental tensor field is rather symmetric as $M\equiv M_{\alpha\beta}$ (equivalently could be both upper indexes), requiring $M_{\alpha\beta}\to M_{\alpha}^{\beta}$ by product with $\xi=\textrm{diag}(-1,1,1,1)$ spacetime signature matrix, e.g. transforming:
\be[M_\mu,M_\nu]\to M_\mu \xi M_\nu- M_\nu \xi M_\mu\qquad\textrm{for}\quad \xi=\textrm{diag}(-1,1,1,1)\ee
For Lagrangian we further took Frobenius norm of such commutator, which for Lorentz invariance becomes:
\be \|X\|^2_F \to \|X\|_\xi^2:= \textrm{Tr}(X\xi X^T \xi)= \sum_{\mu\nu} X_{\mu\nu} X^{\mu\nu} \ee
with also negative contributions, well known e.g. in EM Lagrangian:
$$ \|F\|_\xi^2=\sum_{\mu\nu} F_{\mu\nu}F^{\mu\nu} =2\sum_{1\leq \mu<\nu \leq 3} \left(F_{\mu\nu}\right)^2 -2\sum_{\mu=1}^3 \left(F_{0\mu}\right)^2 $$
Finally Lorentz invariant Lagrangian can be chosen like 4th order term in skyrmion model as (we can use $[\partial_\mu M,\partial_\nu M]=\partial_\mu M\, \xi\, \partial_\nu M-\partial_\nu M\, \xi\, \partial_\mu M$ operating on matrices):
\be \mathcal{L}=-\sum_{\alpha\beta\mu\nu} F_{\mu\nu\alpha\beta} F^{\mu\nu\alpha\beta}-V(M)\ee
$$\textrm{for}\qquad F_{\mu\nu\alpha\beta}=[\partial_\mu M,\partial_\nu M]_{\alpha\beta} $$
with electromagnetic $F$ tensor extended to include also quantum phase governed by Klein-Gordon-like equation, and second set of Maxwell equations for gravity.

Figure \ref{Fmunu} contains practical approximation of $\bar{F}_{\mu\nu}$ tensor as $F_{\mu\nu}$ with reversed rotation and boost. It is for vacuum case with $D=\textrm{diag}(g,1,\delta,0)$ shape for $g\gg 1\gg\delta>0$ formula (\ref{RF}) for $O$ containing rotation + boost and $\Gamma_\mu=O^T O_\mu$. Affine connection, curvature without tilde $\Gamma, R\equiv R^e$ is for spatial rotations, with tilde:x $\tilde{\Gamma}, \tilde{R}\equiv R^g$ for the 0th time axis.

Hamiltonian can be calculated as previously (\ref{hamil}), replacing Frobenius norm with Lorentz invariant one - having both positive spatial energy contributions, but surprisingly also negative $\alpha 0$ energy contributions (as Legendre transform changes only 2 out of 4 indexes of curvature tensor - corresponding to derivation directions, not SO(1,3) generators inside):
\be \mathcal{H}=\sum_{0\leq \mu<\nu\leq 3} F_{\mu\nu\alpha\beta} F_{\mu\nu}^{\alpha\beta} +V(M)\ee
$$ \mathcal{H}=2\sum_{0\leq \mu<\nu\leq 3}\left(\sum_{1\leq\alpha<\beta\leq 3} (F_{\mu\nu\alpha\beta})^2-\sum_{\alpha=1}^3  (F_{\mu\nu\alpha0})^2    \right)+V(M)  $$

Positive contribution spatial part of $\bar{F}_{\mu\nu}$ in Figure \ref{Fmunu} contains sum of spatial curvature $R$: for $R^1_{\mu\nu}=\Gamma^2_\mu \Gamma^3_\nu -\Gamma^3_\mu \Gamma^2_\nu$ corresponding to EM, and $R^2, R^3$ multiplied by tiny $\delta$ corresponding to QM. It is summed with curvature of time axis $R^g \equiv \tilde{R}$ corresponding to GEM. Energy density (Hamiltonian) contains squares of such sums. Gravity fluctuations are usually many orders of magnitude slower, hence expanding such positive energy term into expected values:
$$E[(R +g^2 \tilde{R})^2]= E[R^2] +g^4 E[\tilde{R}^2] +2g^2 E[ R \tilde{R}] $$
we can treat EM (minimizing first term) and gravity (second) as nearly independent, Fig. \ref{source}, \ref{4Deq} derive Maxwell-like equations for EM and GEM. Their dependence in $2g^2 E[R\tilde{R}]$ term makes e.g. that in \textbf{presence of gravitational field $(\tilde{R})$ there is slowing down of EM and QM propagation}, as in \href{https://en.wikipedia.org/wiki/Variable_speed_of_light}{variable speed of light}~\cite{dicke} - leading e.g. to gravitational time dilation, and refractive index for light bending through Fermat principle.

The most interesting are negative energy contributions in Hamiltonian due to spacetime signature (0th row and column of $\overline{F}_{\mu\nu}$), which seem unavoidable as also e.g. in Dirac equation. They are $\Gamma \tilde{\Gamma}$ rotation-boost type, e.g. leading to tendency for spatial rotations (toward time $\Gamma_0$ or space $\Gamma_1,\Gamma_2,\Gamma_3$) in presence of gravity. Especially for twists $\Gamma^1_0$ in $\Gamma^1_0 \tilde{\Gamma}_i$ terms - exactly as required to \textbf{propel electron's de Broglie clock/angular momentum (and neutrino oscillations)} as in Fig. \ref{kink}) by mass itself. This tendency comes together with for nonzero $\tilde{\Gamma}$, what is crucial for \textbf{gravitational mass}: local boosts preferred by presence of particle. 

In contrast, $\Gamma \Gamma$, $\tilde{\Gamma} \tilde{\Gamma}$ same type products have positive energy contributions, leading to energetic preference for only single nonzero $\Gamma$ and $\tilde{\Gamma}$ - prevented e.g. by matter, or various types of noise like cosmic microwave background radiation, and noise of other degrees of freedom - which should be thermalized, but difficult to observe, hence acting as dark energy/matter. Energy levels (temperature) of such noise weakens throughout the history of the Universe, what might be crucial in cosmological models. Generally the  negative energy contributions should have tendency to form  \href{https://en.wikipedia.org/wiki/Void_(astronomy)}{cosmic voids} of locally lowered noise levels, and indeed lots of them are observed.

For Newton force naively we would get the same sign as for Coulomb this way, what needs to be inversed to make same masses attract. These negative energy contributions seem crucial for such inverse, suggesting e.g.  $\Gamma \propto \tilde{\Gamma}$ statistical dependence. Additionally, there are changes of lengths of axes ($M$ eigenvalues) - e.g. for noise as dark matter/energy contribution, also necessary in particles for regularization - crucial for their mass, gravity, oscillations. Such eigenvalue derivatives bring additional EM-like energy contributions, which seems to correspond e.g. to gluons as in Fig. \ref{GEM}: 8 generators of Yang-Mills theory.  

Regarding the Higgs-like potential $V(M)$ minimized for a fixed set of eigenvalues (shape), one approach could be like Landau-de Gennes using traces of powers, now we need to modify it for SO(1,3) to include rotations and boosts. Using $M\equiv M_{\alpha \beta}$, traces of $(M\xi)^p = (-\lambda_0)^p + \lambda_1^p +\lambda_2^p+ \lambda_3^p$ are rotation-boost invariants, we could use e.g. $V(M)=\sum_{p=1}^k (\textrm{Tr}((M\xi)^p)-c_p)^2$ potential. As potentials are often effective, we should search for an even deeper model to derive such potential, e.g. replacing simple abstract ellipsoids represented by $M$, with some more concrete objects.

There might be required additional $\det(M)=$ const (volume preserving) constraint, deforming 0th axis in presence of field regularization in particles. Such deformation of 0th axis seems crucial especially for regularization of black hole central singularity, merging energy of all the particles which fell there.

Choosing the details especially of potential is very difficult, will rather require PDE simulations. The negative energy term should be usually compensated by positive, e.g. due to noises, e.g. in Fig. \ref{kink} allows to only slightly reduce mass of particle by activating oscillations. Obviously the discussed Lagrangian might be incorrect, or simplified e.g. requiring additional terms like 6th order skyrmion term. Finally such $M$ field represents rotations and boosts of some abstract field - searching for a more concrete one could lead to an even deeper model, e.g. to further reduce the number of parameters and derive potential. 

\section{Correspondence with the Standard Model} \label{sec6}
\begin{figure}[t!]
    \centering
        \includegraphics[width=9cm]{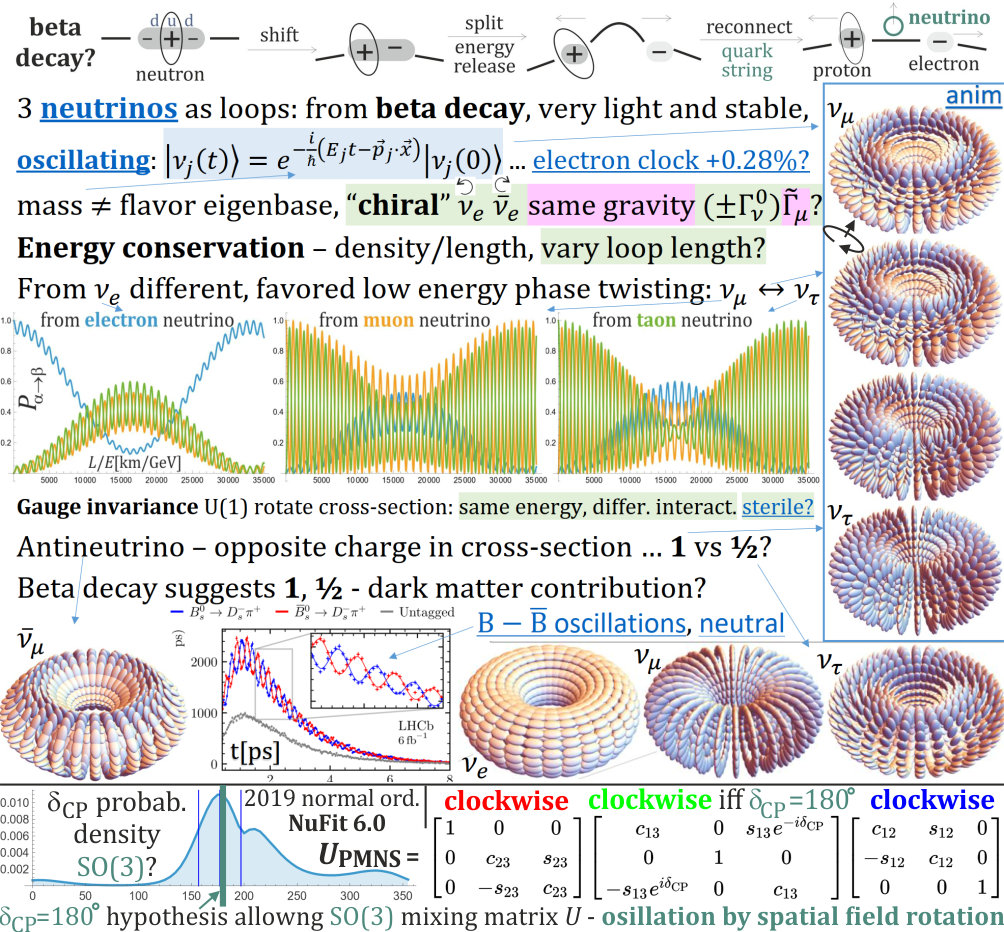}
        \caption{Neutrino model as simple loops of topological vortices - 3 types, very light and stable, rather unavoidable e.g. in quark-gluon plasma called "the most vortical fluid"~\cite{vortical}, string hadronization successful to model LHC collisions, should be created in beta decay (having continuous spectrum).  Size of neutrino wavepacket was recently bounded from below by 6.2pm~\cite{neutsize}: $\sim 1000\times$ larger than nucleus - extremely light vortex loops could achieve. We also need \href{https://en.wikipedia.org/wiki/Neutrino_oscillation}{neutrino oscillations} and  \href{https://en.wikipedia.org/wiki/B\%E2\%80\%93Bbar_oscillation}{B-Bbar oscillation} (diagram source). The oscillation formulas use the same mechanism as for electron's de Broglie clock~\cite{clock}, but with 3 masses of eigenbasis different from flavor, hence leading to oscillations between flavors - as we can see, mainly between muon and taon neutrino, also in the prosed model. To conserve energy, here mass is proportional to length - such loops could also vary length during oscillations. E.g. beta decay has some angle preferences, which might correspond to suggestions of sterile neutrinos. Hamiltonian terms with $\tilde{\Gamma} \Gamma$ seem crucial e.g. for oscillations, and while antineutrinos have the same gravitational mass ($\tilde{\Gamma}$), spatial $\Gamma$ is inverted, suggesting why we have only left-handed neutrinos and right-handed antineutrinos. Shown beta decay mechanism suggests cross-section should have topological charge 1, but 1/2 seem also possible, however, currently lacking creation mechanism - might contribute to dark matter coming e.g. from Big Bang. 
        \textbf{Bottom}: standard PMNS neutrino mixing parametrization becomes SO(3) spatial rotation with 3 clockwise generators for $\delta_{\textrm{CP}}=180^\circ$, which for normal ordering is within $1\sigma$ of NuFit 6.0~\cite{nufit} estimation.}
        \label{neutrinos}
\end{figure}

\begin{figure}[t!]
    \centering
        \includegraphics[width=9cm]{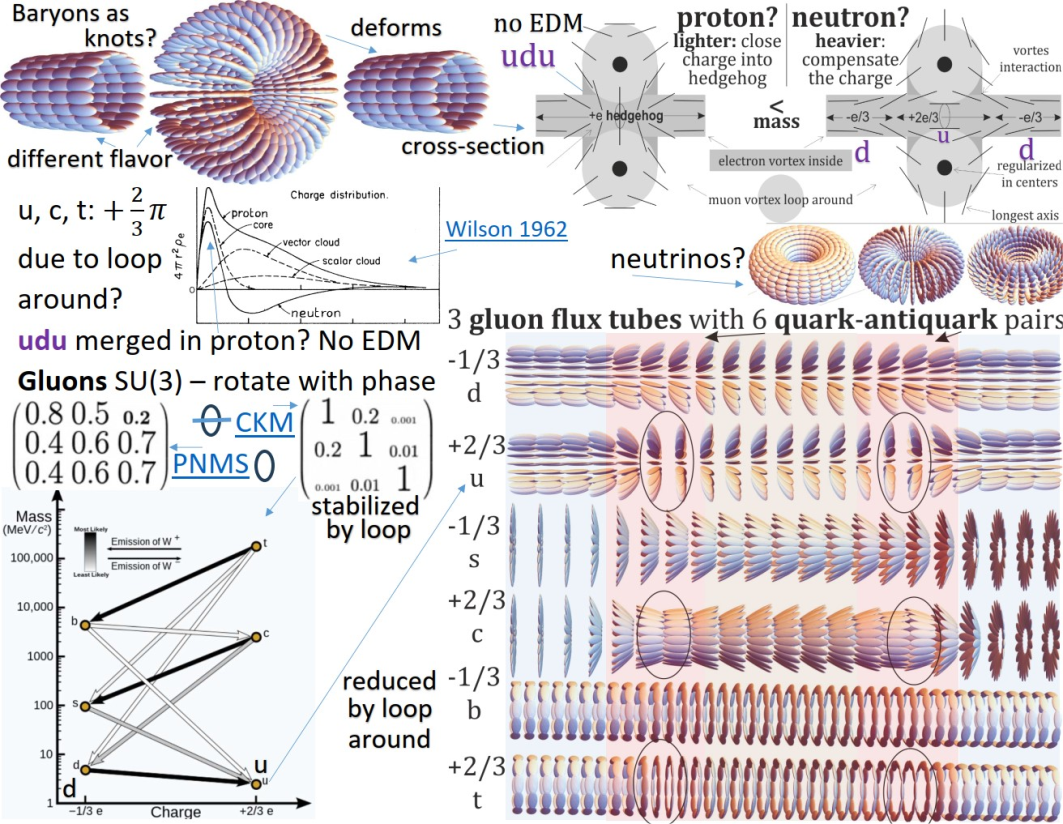}
        \caption{Baryons as the simplest knots, unavoidable e.g. in (topological) string hadronization. As shown, the outer vortex loop enforces charge-like field rotation in the inner vortex, as in Fig. \ref{string} can be fractional charge for quarks. This way baryons structurally require some charge - proton can enclose it into elementary, while neutron has to compensate it - what is costly, explaining why neutron is heavier. Neutron can have 3 quark structure as required - with positive core, negative shell, as in the shown charge distribution plot from Wilson~\cite{wilson}. In proton such 3 quarks seem nearly merged - leading to charge distribution wider than for point charges. There are also shown such 6 quark-antiquark pairs as excitations of topological vortices along one of 3 axes, of 2/3e charge enforced by loop around, this way reducing energy especially of 'u' up quark, and 1/3e charge required to compensate them (energy diagram from \href{https://en.wikipedia.org/wiki/Cabibbo\%E2\%80\%93Kobayashi\%E2\%80\%93Maskawa_matrix}{Cabibbo-Kobayashi-Maskawa Wikipedia article}). There are also shown approximate quark (CKM) and neutrino (PNMS) mixing matrices - the first one is much more diagonal, what can be understood here that internal field rotations for knots (as baryon) are more constrained than for simple loops (as neutrinos) - due to constraining outer loop.   }
        \label{baryon}
\end{figure}

Let us briefly discuss further correspondence between topological excitations of SO(3) vacuum field and particle menagerie of Standard Model, summarized in Fig. \ref{neutrinos}, \ref{baryon}, \ref{part}.
\subsection{Topological string hadronization correspondence}
Searching for such correspondence, the best hint seems \href{http://www.scholarpedia.org/article/Parton_shower_Monte_Carlo_event_generators\#String_model}{string hadronization} of quark string as Abrikosov vortex like in Fig. \ref{string} - a hot string created in high energy particle collisions e.g. in LHC, decays through reconnections into particle shower. Also quark-gluon plasma is referred as "the most vortical fluid"~\cite{vortical}, and such vortices should form e.g. knots. Particles as knots were proposed e.g. as glueballs (\cite{glue1,glue2}) and other exotic objects (e.g. \cite{parknot}), but in LHC collisions more common particles are dominating, so should participate in such correspondence.

Thinking about decay possibilities for topological vortices, and searching for correspondence with such particle shower:

\begin{itemize}
  \item Such vortex can form \textbf{simple loops}, which should be very light and stable - the only particle created in colliders agreeing with this description seems neutrino.
  \item Such vortices could create stable \textbf{knots} - they might be of various sizes, and the only known such varying size objects in particle physics are nuclei, with baryons as the simplest knot: vortex loop around another vortex.
  \item Loops could have internal twist like M\"{o}bius strip - such \textbf{twisted loops} should be statistically quite frequent in collisions, and pions, kaons are dominating there - suggesting to interpret this twist as related to strangeness, also for strange baryons twisting their loop.   
 \end{itemize}

 The above correspondence seems quite constrained - it is a valuable exercise to search for different possibilities, to realize lack of freedom for its modification. Fortunately, as discussed further, at least qualitatively its consequences seem to agree with experimental view on the Standard Model.

For example here are some observed \href{https://en.wikipedia.org/wiki/List_of_baryons}{baryon decay modes} - literally releasing some strangeness by pion, or twice larger by kaon - like releasing part of vortex twist to reduce tension as "strangeness decay" in Fig. \ref{part}: 
$$\textrm{strangeness 3 to 2 or 1:}\quad \Omega^- \to \Xi+\pi\quad \textrm{or}\quad \Lambda^0+K^-$$
$$\textrm{strangeness 2 to 1:}\quad \Xi^- \to \Lambda^0+\pi^-,\quad \Xi^0 \to \Lambda^0+\pi^0$$
$$\textrm{strangeness 1 to 0:}\quad \Lambda^0 \to p^+ + \pi^- \quad \textrm{or}\quad n^0 + \pi^0 $$

Obviously there are hundreds of observed particles, for which we should search for correspondence e.g. with local energy minima in field configuration space. As they are often very short-lived, these could be very shallow minima. Also some are interpretations of perturbative approximation, e.g. Coulomb interaction represented with photon exchange, this way e.g. W, Z bosons could represent dynamical vortex excitations.

\begin{figure}[t!]
    \centering
        \includegraphics[width=9cm]{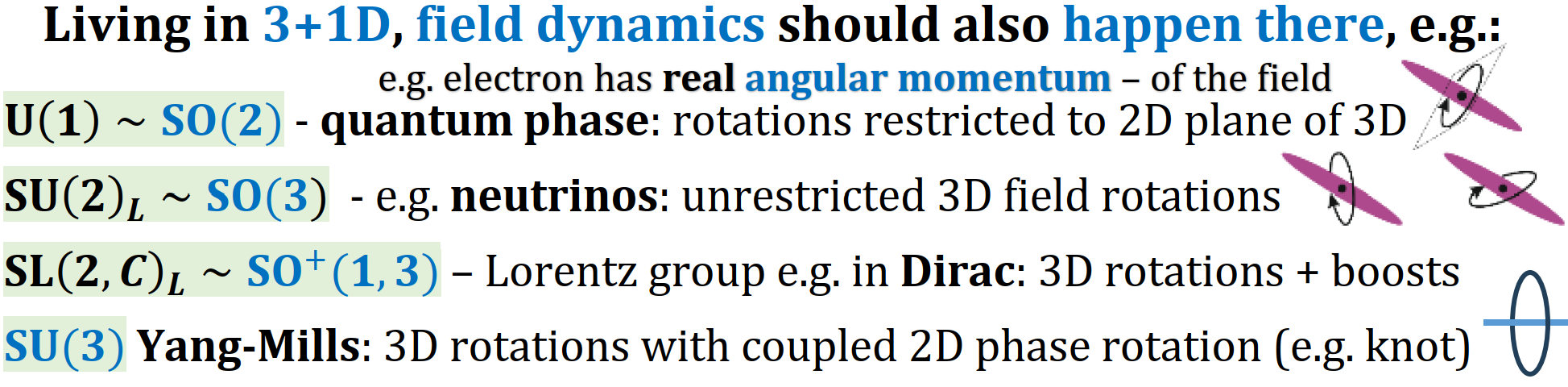}
        \caption{As we live in 3+1D, natural symmetry of the field is SO(1,3) Lorentz group e.g. of \href{https://en.wikipedia.org/wiki/Bispinor}{bispinor} in Dirac equation, and many used can be viewed as abstract representations of its subgroups, or effective extensions. Quantum phase is represented by complex numbers, but e.g. in electron also requires angular momentum, which in field theory means rotation of field in perpendicular plane. Neutrino oscillation $\textrm{SU}(2)_L\sim \textrm{SO}(3)$ can be also viewed as 3D field rotations, but this time not restricted to a plane. Finally quarks require larger SU(3) group, but they are composite objects e.g. knots of quark string here, so this seems effective description of e.g. SO(3) spatial field rotation of one part, requiring coupled phase rotation of second part. }
        \label{group}
\end{figure}

\subsection{Quarks, strings, baryons, nuclei}
Gauss law for a region returns electric charge inside, which should be integer multiplicity of elementary charge $e$, here enforced by making it topological. However, deep inelastic scattering has shown fractional charges inside baryons - which are believed to be connected by 1D quark strings (what also initiated string theory), nonperturbatively modelled as topological Abrikosov-like vortex~\cite{string} - in which there can be added fractional charge excitations through inward/outward field rotation as in Fig. \ref{string}. Gauss law for a region cutting such string has (regularized) singularity/conflict in this point, leading to additional energy density per length, which for agreement with QCD should be asymptotically $\sim 1$ GeV/fm.

Existence of vacuum 1D topological defects as quark string should also have consequences in larger scale, and seems they indeed have. For example in Sun's corona there are seen bright stable lines, called magnetic flux tubes, and suspected to have such topological nature, from \cite{flux}: "Vortices in superfluid Helium and superconductors, magnetic flux tubes in solar atmosphere and space, filamentation process in biology and chemistry have probably a common ground, which is to be yet established. One conclusion can be made for sure: formation of filamentary structures in nature is energetically favorable and fundamental process". Also 1D structures are postulated in cosmological scale as \href{https://en.wikipedia.org/wiki/Cosmic_string}{cosmic strings}, with first claims of experimental confirmation~\cite{cosmic}.

With suggested \textbf{baryon} as the simplest knot, looking at diagram e.g. in \ref{baryon}, we can see that the loop around deforms structure of internal vortex - exactly into  inward/outward field rotation required for charge. If this rotation would be by $\pi$, we would get hedgehog corresponding to elementary charge. Here it can be smaller - requiring a conflict on such string, having additional energy density per length - should be $\sim$ 1 GeV/fm for quark string, making it energetically expensive to take quarks apart - confining them. However, its contribution weakens in high energy for asymptotic freedom, finally leading to widely used \href{https://en.wikipedia.org/wiki/Cornell_potential}{QCD Cornell potential}: first Coulomb interaction between charges, then asymptotically $\sim 1$ GeV/fm linear energy density per length.

\textbf{Proton} can enclose this structurally enforced fractional charge into hedgehog being elementary charge, while \textbf{neutron} has to compensate it to zero total charge - explaining larger mass. It agrees with required charge distributions from literatures e.g. ~\cite{wilson,neutron1} - neutron as having positive core and negative shell, proton only positive but smeared in comparison to point charge - suggesting 3 quarks very close together. 

For \textbf{deuteron}, both baryons require fractional positive charge - can share a single elementary charge to reduce total energy for binding, as in Fig. \ref{part}. This way it is "+-+" charge distribution - has relatively   \href{https://en.wikipedia.org/wiki/Deuterium\#Magnetic_and_electric_multipoles}{ large electric quadrupole moment}, experimentally 0.2859 e\,$\textrm{fm}^2$. In contrast, deuteron as just proton + neutron would have no electric quadrupole moment - requiring e.g. shift of quarks. Deformation of quark distribution when nucleons combine into nuclues is generally referred as \href{https://en.wikipedia.org/wiki/EMC_effect}{EMC effect}~\cite{EMC}, and seems not currently understood - the proposed model could help with. Also spins of both baryons should be aligned as $\mu_d\approx \mu_p +\mu_n$, what agrees with predicted view.

Quark strings are mostly considered inside nucleons, but could also go out and contribute in \textbf{nuclear binding}, helping to overcome Coulomb repulsion of protons. Especially for \href{https://en.wikipedia.org/wiki/Halo_nucleus}{halo nuclei}~\cite{halo} - for milliseconds binding single (halo) neutrons or protons in a few femtometer distance, much larger than $\sim 1$ fm of standard \href{https://en.wikipedia.org/wiki/Nuclear_force}{nuclear force}. Believed to require 3-body forces~\cite{halo3}, forming Borromean structures. For example Lithium-11 binding 2 neutrons for $\approx 8.6$ milliseconds in a few femotometer distance, larger than of strong force. Binding them through connections by 1D quark strings could explain both stability in large distance, and 3-body forces required in effective description. Such models of nuclei as knots of quark strings might be very useful to optimize nuclear processes like fusion.

As there is no Gauss law for baryon number, many models allow its violations, including such view on baryon as knots of quark strings. Such violation is required e.g. for \href{https://en.wikipedia.org/wiki/Baryogenesis}{baryogenesis} (creation of more baryons after Big Bang), or (massless) \href{https://en.wikipedia.org/wiki/Hawking_radiation}{Hawking radiation}. The latter suggests ultimate energy source by squeezing baryons into black hole, and gathering energy from its evaporation. If possible through black hole, it should be doable also directly by stimulation of e.g. proton decay, for example optimizing parameters for particle collisions, or shooting incoming proton beam with free electron laser. For baryon as knot e.g. trying to swing to untangle: internal charge of proton with electric field, or twist the magnetic dipole with magnetic - finalizing the proposed model, we could numerically optimize its parameters.

\subsection{Neutrinos as loops of quark strings}
While EM waves are stopped by a centimeter of lead, neutrinos can easily pass through the Earth - need some additional stabilization mechanism, like topological. Their basic creation mechanism is beta decay of neutron to proton - releasing energy not only as photons, but also as such very stable neutrinos.

From the other side, having quark strings it seems unavoidable for them to form loops, which should be interpreted as particles, would be very light and stable - and we know only neutrinos in agreement with these properties. Recently size of neutrino wavepacket was experimentally bounded from below by $\approx 6.2$pm, what is a thousand times larger than nucleus - we need to understand their field configurations hidden there, and very light vortex loop might achieve such large sizes.

Also they have 3 flavors - as topological vortices: along one of 3 axes as we live in 3D. They can oscillate between flavors, vortices through field rotations, like in Fig. \ref{neutrinos}, using also $\textrm{SU}(2)_L\sim SO(3)$ rotation group in 3D. Oscillation for low energy twist should be more likely - and indeed there is dominating between muon and taon neutrino. As expected, for electron neutrino it is essentially different.

These oscillations should be propelled by the mass itself, as for electron's clock using $\psi\propto \exp(-iEt/\hbar)$ phase evolution for $E=mc^2$ relativistic mass, also angular momentum (of the field not point particle). These configurations are relatively simple - are planned to be studied in details to understand this propulsion through mass. Field rotations (affine connection) are: around vortex, around loop, temporal for oscillations $\Gamma_0$, and also boosts $\tilde{\Gamma}$ from mass - products with boosts have negative energy contributions in Hamiltonian in Fig. \ref{Fmunu}, however, time dependence has also positive energy contributions in $\Gamma_0 \Gamma_i$ terms, hence activation of oscillation should allow only small reduction of particle mass for finite frequency, like in Fig. \ref{kink}. Energy minimization should lead to the observed oscillation parameters, allowing to constrain the model. For antineutrino, gravitational mass should be the same, but field rotation around vortex is reversed - what should explain left/right-handedness of (anti)neutrinos.

As in Fig. \ref{group}, while electron has only U(1) degree of freedom for field rotations which seems corresponding to angular momentum, for vortices we have freedom for field rotation SO(3)$\sim$ SU$(2)_L$ (as SU(2) is double covering of SO(3)). Direct electron clock experimental confirmation~\cite{clock} required $0.28\%$ higher energy than predicted - this difference might come from still 3 types of tendencies (as for neutrino) acting in electron, but kind of being projected (added) into single allowed evolution degree of freedom.

Finally to effectively describe weak interactions, we indeed need U(1)$\times$SU$(2)_L$ symmetry group. Such field rotations for vortices are relatively unconstrained, in contrast to their knots - with loops blocking field rotations, hence making mixing matrices much more diagonal (\href{https://en.wikipedia.org/wiki/Cabibbo\%E2\%80\%93Kobayashi\%E2\%80\%93Maskawa_matrix}{CKM} vs \href{https://en.wikipedia.org/wiki/Pontecorvo\%E2\%80\%93Maki\%E2\%80\%93Nakagawa\%E2\%80\%93Sakata_matrix}{PNMS}), enforcing additional phase change during spatial rotations - requiring to use full SU(3) symmetry group for strong interactions. As mentioned in Fig. \ref{GEM}, gluons in Yang-Mills term could be also interpreted as field curvature with one connection up the Higgs potential: changing $M$ eigenvalue. 

To oscillate changing flavors, there are needed 3 different masses - suggesting that oscillations can change mass, what naively would mean violation of energy conservation. It is repaired here as this is mass per length - allowing to conserve energy by also varying loop length. A basic source of neutrino is beta decay of neutron, which has continuous energy spectrum, suggesting also varying length of created vortex loop, mass of the electron antineutrino would be proportional to this length. While experimental data finds lighter and lighter neutrinos, maybe we should not treat them as lower boundaries of the mass - this mass could be varying (here with loop length), as for continuous spectrum we should rather consider neutrino mass probability distribution.

Cross-section of such vortex loop has topological charge - beta decay suggests it should be $\pm 1$. However, spin 1/2 suggests field with $n \equiv -n$ type symmetry, allowing also for vortices with $\pm 1/2$ charge in cross-section, which would be even more difficult to create and interact with - formed e.g. in Big Bang might contribute to dark matter. Additionally, neutrino creation processes like beta decay seem to have additional constraints for angles here, what should also concern their interaction - looking like additional mechanism violating assumed group symmetries, which might offer alternative solution for problems leading to introduction of \href{https://en.wikipedia.org/wiki/Sterile_neutrino}{sterile neutrinos} - not by additional separate particles, only less likely interacting intrinsic states of standard neutrinos.

\subsection{Hints for issues of Standard Model}
The goal of the discussed approach is not to replace the Standard Model, only searching for more fundamental nonperturbative model - without effective: probability distributions and abstract creation operators, replacing the latter with concrete field configurations of particles, leading to probabilities by Feynman averaging. It should allow to reduce the number of paraments, e.g. calculating masses by integrating Hamiltonian. Also for better understanding, e.g. providing answers where Standard Model has issues, for example:
\begin{itemize}
  \item Standard Model has originally predicted zero neutrino masses, which turned out nonzero - as vortex loops here, they have mass/energy density per length.
  \item Quark mass is only about 1\% of proton mass - here baryon is much larger structure, with mass mostly in field deformations which can be interpreted as gluons.
  \item \href{https://en.wikipedia.org/wiki/Proton_spin_crisis}{Proton spin crisis}~\cite{spincrisis} - while it was expected that quarks carry all proton's spin, experimental data suggest it is barely $4-24\%$. For proton as simple knot, angular momentum is distributed over the entire field configuration,
  \item \href{https://en.wikipedia.org/wiki/Proton_radius_puzzle}{Proton radius puzzle} - (root mean squared) radius through interactions with electron measured as $\approx 0.877$ fm, with muon $\approx 0.842$ fm~\cite{protonsize}. Here electron and muon should differ by $\pi/2$ field rotation, what should modify interactions with proton, also e.g. using taon instead.
  \item \href{https://en.wikipedia.org/wiki/Free_neutron_decay\#Neutron_lifetime_puzzle}{Neutron lifetime puzzle} - turns out that cold "bottle" neutrons have $\approx 878$ seconds lifetime, while hot "beam" $\approx 887$ seconds~\cite{neutronlife, exciting} - neutron as knot has intrinsic excitation modes, which deexcitation time should correspond to this time difference.
  \item \href{https://en.wikipedia.org/wiki/EMC_effect}{EMC effect} - turned out quark distributions in nucleons are modified when binding into nucleus~\cite{EMC} - also here leading e.g. to electric quadrupole moment of deuteron.
  \item \href{https://en.wikipedia.org/wiki/Three-body_force}{Three-body force}, \href{https://en.wikipedia.org/wiki/Halo_nucleus}{halo nuclei} seem not well understood, here rather requiring quark strings helping with nuclear binding - including  neutron/proton halos, and in effective description requiring 3-body interactions. 
  \item For dark matter e.g. lighter versions of neutrinos in Fig. \ref{neutrinos} are suggested. While CMBR is $\approx 2.7$K EM noise, degrees of freedom of other interactions should also thermalize, being very difficult to directly observe - interpreted as dakr energy. Baryons could be created in baryogenesis here. 
  \item While we get particle menagerie from SO(3) field dynamics, extending to SO(1,3) we automatically get unification with gravity as GEM. As observationally \href{https://en.wikipedia.org/wiki/Shape_of_the_universe}{spacetime seems flat}, here only space is curved in flat 4D. 
\end{itemize}

While these seem valuable suggestions for issues of Standard Model, the details will rather require finalizing such proposed deeper model and performing numerical simulations.

\section{Conclusions and further work}
There was presented LdGS mathematical framework combining Landau-de Gennes field with Skyrme kinetic term, which field dynamics unifies EM (Maxwell form high energy twists) + QM (Klein-Gordon from low energy twists) + GEM (Maxwell from very high energy boosts). Its topological excitations start with point-like charges as quantized electric, and vortices as quark strings - building particles e.g. by string hadronization, getting promising agreement with Standard Model.

This article is work in progress, planned to be further developed: aiming as good agreement with particle physics as possible - both for better understanding, also maybe to try to recreate some phenomena e.g. in liquid crystal experiments, like observation of additional fluxon-like vortex (disclination) for hedgehog configuration in biaxial nematic (no naked charges), transformation between 3 types of vortices as neutrino oscillation analogy, their quark-like excitations, hadronization as decay of such vortex into particles, etc.

The main open question is choosing the potential - e.g. depending only on the shape $V(M)$, or maybe also derivatives $V(A)$, ideally with minimal number of parameters, like only $g$ and $\delta$ preferred eigenvalues, or less as e.g. $g\approx 1/\delta$. As potentials are usually effective, the best would be deriving it from a deeper model, e.g. replacing abstract ellipsoids with concrete objects. A natural way to find the potential is through search for agreement e.g. with 3 leptons as hedgehog of one of 3 axes as in Fig. \ref{intr}: they should form 3 local minima in the space of possible rotations of hedgehog ansatz, probably stabilized by the enforced topological vortices.

We should also get neutrino oscillations enforced by mass, and analogous electron's intrinsic periodic process (Zitterbewegung/de Broglie clock~\cite{clock}) probably due to negative energy terms in Hamiltonian as in Fig. \ref{Fmunu}. Details are yet to be developed, e.g. gravitational mass might require e.g. fixing $\det(M)$ constraint suggested in \cite{my}. EM-GEM interaction slowing down EM propagation should explain graviational time dilation and lensing.

Calculations like started in Fig. \ref{coulomb}, \ref{newton}, \ref{source} and \ref{4Deq} are crucial direction of development, also parametrizations to moduli space, trying to extend correspondence with particle physics, finally performing 2nd quantization, Feynman ensembles aiming agreement with the Standard Model. Preferably also full hydrodynamical simulations to better understand particle configurations and dynamics.

\begin{figure*}[b!]
    \centering
        \includegraphics[width=18cm]{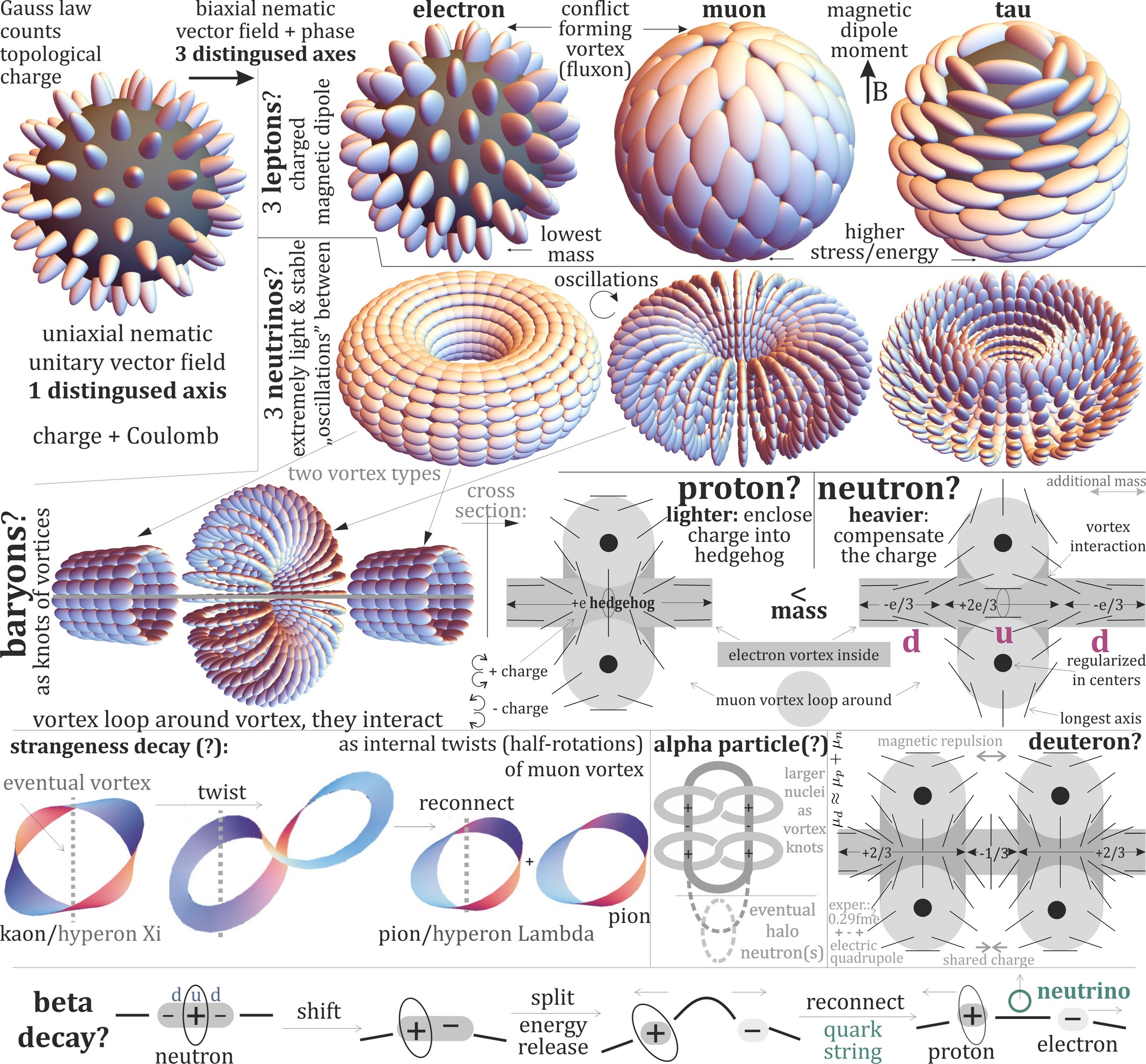}
        \caption{Further particle-like configurations nearly required by string hadronization as topological - discussed in Section \ref{sec6}. Hedgehog configurations of one of 3 axes resemble \textbf{3 leptons}: the same electric charge (as topological), but different realization: regularization, mass. Hairy ball theorem says there is a conflict of axes on the sphere - leading to outgoing low energy vortex (2D topological charge) of one of 3 types (along one of 3 axes) - like fluxons in superconductor carrying magnetic field: enforcing magnetic dipole moment for leptons. Loop of such vortex would be very light and difficult to interact with, resembling \textbf{3 neutrinos} - with possible oscillations between each other through field rotation (Fig. \ref{neutrinos}), they should be produced in beta decay. Such vortices, corresponding to "quark strings" in QCD, can further form knots, which resemble baryons, nuclei. As in Fig. \ref{baryon}, interaction between vortices inside such knot with vortex loop around enforces charge-like (hedgehog) configuration inside: makes that \textbf{baryon configuration requires some charge} - can be fractional, but all sum to integer charge (confinement). Proton can just enclose this charge (hedgehog), but neutron has to compensate it to zero - suggesting \textbf{why neutron is heavier} (naively should be lighter due to charge), through quark-like fractional charge structure. Such concluded: positive core, negative shell charge distribution of neutron is suggested in literature e.g. \cite{wilson, neutron1}. It also suggest \textbf{binding mechanism for deuteron}: as two baryons sharing required charge - also explaining observed relatively large electric quadrupole moment (experimentally: 0.2859 e\,$\textrm{fm}^2$) and aligned spins $(\mu_d\approx \mu_p +\mu_n)$. Binding of larger nuclei could additionally use vortices forming stable knots, e.g. \textbf{halo nuclei} with neutrons stably bind in large a few femtometer distance, and effectively require 3-body forces. Vortex loop might have additional internal twist, which is quantized ($k\pi/2$) and resembles \textbf{strangeness} - relaxed through muon/kaon production in decay of strange baryons. }
        \label{part}
\end{figure*}

To summarize, while currently the main focus is on QFT pertubative approximations, the real situation is given by nonperturbative QFT - understanding field configurations of particles and Feynman diagrams, before considering their Feynman ensembles. Especially the string hadronization as topological requires quite constrained correspondence, discussed in Section \ref{sec6} and summarized in Fig. \ref{part}. While we automatically get looking perfect qualitative agreement, quantitative will require further work: finalizing the model including potential, and performing (numerical) calculations - e.g. to derive $\approx 30$ parameters of Standard Model  from $\approx 2$: $\delta\sim 10^{-10}$ for QM and $g\sim 10^{10}$ for gravity energy scales. Finally, the discussed $M$ is some abstract field recognizing SO(1,3) dynamics - searching its concrete realization could  lead to even deeper model, e.g. to further reduce the number of parameters.

\bibliographystyle{IEEEtran}
\bibliography{cites}

\end{document}